%% file: 0main.tex
\documentclass[manuscript,screen]{acmart}
\input{00packages.tex}

\AtBeginDocument{%
  \providecommand\BibTeX{{%
    \normalfont B\kern-0.5em{\scshape i\kern-0.25em b}\kern-0.8em\TeX}}}

\setcopyright{none}
\copyrightyear{2024}
\acmYear{2024}
\acmDOI{XXXXXXX.XXXXXXX}

\acmConference[Arxiv]{}{2024}{}
\acmBooktitle{Arxiv} 
\acmPrice{15.00}
\acmISBN{978-1-4503-XXXX-X/18/06}

\AtBeginShipoutNext{\AtBeginShipoutUpperLeft{%
  \put(\dimexpr\paperwidth-13cm\relax,-1.5cm){\makebox[0pt][r]{\large \textcolor{blue}{Submitted for publication in January 2024}}}%
}}

\begin{document}

\title{Navigating AI Fallibility: Examining People's Reactions and Perceptions of AI after Encountering Personality Misrepresentations}
\renewcommand{\shorttitle}{Navigating AI Fallibility}


\author{Qiaosi Wang}
\affiliation{%
  \institution{Georgia Institute of Technology}
  \city{Atlanta, GA}
  \country{USA}}
\email{qswang@gatech.edu}

\author{Chidimma L. Anyi}
\affiliation{%
  \institution{Georgia Institute of Technology}
  \city{Atlanta, GA}
  \country{USA}}
\email{canyi3@gatech.edu}

\author{Vedant Das Swain}
\affiliation{%
  \institution{Northeastern University}
  \city{Boston, MA}
  \country{USA}}
\email{v.dasswain@northeastern.edu}

\author{Ashok K. Goel}
\affiliation{%
  \institution{Georgia Institute of Technology}
  \city{Atlanta, GA}
  \country{USA}}
\email{ashok.goel@cc.gatech.edu}







\renewcommand{\shortauthors}{Wang et al.}

\begin{abstract}
  Many hyper-personalized AI systems profile people’s characteristics (e.g., personality traits) to provide personalized recommendations. These systems are increasingly used to facilitate interactions among people, such as providing teammate recommendations. Despite improved accuracy, such systems are not immune to errors when making inferences about people’s most personal traits. These errors manifested as~\textit{AI misrepresentations}. However, the repercussions of such AI misrepresentations are unclear, especially on people’s reactions and perceptions of the AI. We present two studies to examine how people react and perceive the AI after encountering personality misrepresentations in AI-facilitated team matching in a higher education context. Through semi-structured interviews (n=20) and a survey experiment (n=198), we pinpoint how people’s existing and newly acquired AI knowledge could shape their perceptions and reactions of the AI after encountering AI misrepresentations. Specifically, we identified three rationales that people adopted through knowledge acquired from AI (mis)representations: AI works like a machine, human, and/or magic. These rationales are highly connected to people’s reactions of over-trusting, rationalizing, and forgiving of AI misrepresentations. Finally, we found that people’s existing AI knowledge, i.e., AI literacy, could moderate people’s changes in their trust in AI after encountering AI misrepresentations, but not changes in people’s social perceptions of AI. We discuss the role of people’s AI knowledge when facing AI fallibility and implications for designing responsible mitigation and repair strategies. 
\end{abstract}

\ccsdesc[500]{Human-centered computing~Empirical studies in HCI}
\ccsdesc[500]{Computing methodologies~Intelligent agents}

\keywords{automated personality recognition, wizard of oz, AI literacy, theory of mind, mental model}



\maketitle

\input{1introduction}

\input{2relatedwork}

\input{3studyoverview}
\input{4study1}

\input{5study2}

\input{6discussion}

\input{7conclusion}
\bibliographystyle{ACM-Reference-Format}
\bibliography{reference}

\appendix
\input{8appendix}





\end{document}
\endinput

%% file: 00packages.tex
\usepackage{amsmath}
\usepackage{multirow}
\usepackage{pbox}
\usepackage{caption}
\usepackage{subcaption}
\usepackage{atbegshi,picture}

\definecolor{DarkBlue}{HTML}{00008B}
\definecolor{green}{rgb}{0.0, 0.65, 0.31}
\definecolor{purple}{HTML}{ae017e}

\newif{\ifhidecomments}
  \hidecommentsfalse 
\ifhidecomments
    \newcommand{\chelsea}[1]{}
    \newcommand{\chidy}[1]{}
    \newcommand{\vedant}[1]{}
    \newcommand{\ashok}[1]{}

\else
    \newcommand{\chelsea}[1]{\textbf{\small\sffamily{\textcolor{purple}{[#1 -- Chelsea]}}}}
    \newcommand{\vedant}[1]{\textbf{\small\sffamily{\textcolor{DarkBlue}{[#1 -- Vedant]}}}}
    \newcommand{\chidy}[1]{\textbf{\small\sffamily{\textcolor{green}{[#1 -- Chidy]}}}}
    \newcommand{\ashok}[1]{\textbf{\small\sffamily{\textcolor{red}{[#1 -- Ashok]}}}}
  \fi

%% file: 1introduction.tex
\section{Introduction}
Recently, a plethora of hyper-personalized AI systems that can profile users' characteristics and traits have been deployed in people's daily lives, with the ultimate goal of providing personalized shopping, music, and social media recommendations. As these systems become more advanced in profiling people's most personal and complex traits such as personalities and emotions~\cite{hall2017say,gou2014knowme,liao2021crystal}, they sometimes give people the illusion that ``machines can read our minds''~\cite{guzman2020ontological}. This illusion has led to various---rather concerning---reactions and perceptions of AI with people attributing AI with beyond-human expertise at reading people's emotions and personalities~\cite{warshaw2015can,hollis2018being}. However, people's perceptions and reactions of AI when this illusion is broken in the face of~\textit{AI misrepresentations} have not yet been explored.

AI misrepresentations is one type of AI fallibilities when AI misinterprets people's most intimate and complex traits like personality and emotions, aspects where people possess the most self-awareness. Prior work has suggested that AI mistakes on human-AI tasks could erode people's trust and social perceptions (e.g., anthropomorphism, intelligence, likeability) of the AI~\cite{honig2018understanding,salem2015would}. However, when faced with AI misrepresentations of people's most intimate traits, people may dismiss it based on their self-awareness, resulting in lack of adherence and trust in AI; or people might exhibit unwavering trust in AI, allowing it to persuade them into accepting false information about themselves.

Even algorithms with supposedly high accuracy can make mistakes when powering hyper-personalized AI systems in-the-wild~\cite{rao2015they}. Understanding people's reactions and perceptions of the AI after encountering AI misrepresentations could offer valuable insights into whether and how people changed their intuitions, beliefs, and reactions of AI in the face of AI fallibilities. This could provide critical implications for the future design and development of responsible interventions, mitigation, and repair strategies to retain user trust, minimize harms, and prevent overreliance when such AI systems inevitably err. In this paper, we seek to understand people's reactions and perceptions of AI after encountering personality misrepresentations by AI. We contextualized this research aim in higher education given the increasing use of hyper-personalized AI systems to identify student characteristics and traits, especially personality traits, to facilitate school project team formations~\cite{xiao2019should,alberola2016artificial,jahanbakhsh2017you,lykourentzou2016personality}. Specifically, this paper explores three research questions:


\begin{table}[h]
\centering
\vspace{-0.5em}
\begin{tabular}{@{}l@{}p{0.92\columnwidth}@{}}
    \textbf{RQ 1: } & What perceptions and reactions do students have about the AI after encountering AI misrepresentations of their personality traits in AI-facilitated team matching?\\
    \textbf{RQ 2: } & How do students change their perceptions of the AI after encountering AI misrepresentations of their personality traits in AI-facilitated team matching? \\
    \textbf{RQ 3: } & What factors contribute to students' perception changes after encountering AI misrepresentations of their personality traits in AI-facilitated team matching? \newline 
\end{tabular}
\vspace{-0.9em}  
\end{table}

\vspace{-0.7em}
To answer these research questions, we conducted semi-structured interviews with twenty college students (Study 1) and a large survey experiment (Study 2) with 198 students on the Prolific platform. In both studies, we took a Wizard-of-Oz approach to fabricate intentionally inaccurate/accurate personality inferences based on participants' personality ground truth. We showed participants in both studies their ``AI-generated personality inferences'' to elicit their perceptions and reactions of AI misrepresentations. We found that people's existing and newly acquired AI knowledge plays a critical role in shaping their perceptions and reactions after encountering AI misrepresentations. Specifically, we pinpointed three rationales that people adopted through knowledge acquired from AI (mis)representations: AI works like a machine, human, and/or magic. These rationales are highly connected to their reactions of over-trusting, rationalizing, and forgiving of AI misrepresentations. We also found that people's existing AI knowledge, i.e., AI literacy, significantly moderate the level of changes in people's overall trust after encountering AI misrepresentations. 



The contribution of our work is three-fold. First, we highlight the role of people's existing and newly acquired AI knowledge in shaping their perceptions and reactions to AI after encountering AI misrepresentations. Second, we provide a descriptive account of people's re-framing of their perceptions through adopting various rationales in the face of AI misrepresentations.
We connect this process to people's reactions of over-trusting, rationalizing, and forgiving of AI misrepresentations. Finally, we provide a set of implications and future directions for the design and development of intervention, mitigation, and repair strategies to consider people's AI knowledge to reduce potential harm. 



%% file: 2relatedwork.tex
\section{Related Work}

\subsection{People's Perceptions and Reactions of Hyper-Personalized AI}
Much HCI work has been conducted to understand people's reactions and concerns to hyper-personalized AI systems that can profile people's emotions~\cite{wang2020sensing,disalvo2022reading,hollis2018being} and personality characteristics~\cite{gou2014knowme,warshaw2015can,liao2021crystal,volkel2020trick}. Prior work found that most people perceived their AI-generated personality profiles to be ``creepily accurate''~\cite{gou2014knowme,warshaw2015can,liao2021crystal,kim2020understanding}. Other studies demonstrated people's tendency to over-trust AI-generated personality profiles about them. Studies showed that people felt unqualified to modify their personality profile generated by the ``expert'' algorithm~\cite{warshaw2015can}, sometimes even overriding their own personal judgments about themselves due to the belief that the AI algorithm could identify their ``hidden self'' and had privileged information about them~\cite{hollis2018being}.

Scholars have developed many theories to explain people's reactions and perceptions of AI systems~\cite{sundar2019machine,sundar2020rise,devito2017platforms,nass1994computers,guzman2020ontological}. Theories such as Machine Heuristic, a rule of thumb that people believe machines are logical, objective, and emotionless, and hence more trustworthy than humans, has been used to explain people's tendency to over-trust AI outcomes~\cite{oh2022you,banas2022machine,kapania2022because,edwards2016robots}; the Computers Are Social Actors (CASA) paradigm has been used to explain people's social reactions to forgive, tolerate, and justify AI misfires~\cite{nass1994computers,kapania2022because}; learning science theories such as conceptual changes~\cite{ozdemir2007overview} and ontological shift~\cite{slotta2011defense,chi1994things} have also been used to examine how people conceptualize the ontological differences between humans and computers~\cite{guzman2020ontological}. Together, these theories suggest people's increasingly blurred conceptualizations and reactions between machines and humans due to technologies' human-like behaviors and capabilities~\cite{guzman2020ontological}. 

Empirical studies in HCI have examined people's various folk theories~\cite{devito2017platforms,liao2021crystal}, mental models~\cite{gero2020mental,wang2021towards}, and factors that influence these perceptions of AI systems. Scholars have found that people with limited folk theories about the data source, data scope, and personalization of the AI algorithm tend to dismiss AI's personality inferences as less threatening~\cite{liao2021crystal}. Additionally, factors such as the data-driven aspect of the system, the ambiguity around how the system works could all influence people's trust and perceptions of the result of AI-generated personality inferences~\cite{kim2020understanding}. Research has also found that people's AI literacy~\cite{druga2017hey,long2020ai,shin2022people,schoeffer2022there} and personal investment in the AI's output~\cite{eslami2019user} play a crucial role in shaping their perceptions, especially trust, in the AI systems. This suggests that people's pre-existing and evolving knowledge about AI could be critical in their perceptions of the AI, which has been under-explored. 

\subsection{Responding and Mitigating AI Fallibilities} Although much work in human-robot interaction has examined people's perceptions of AI agents after encountering AI failures, these studies suggest mixed results. On the one hand, prior work showed that encountering robot failures would significantly lower people's perceptions of the robot~\cite{honig2018understanding} in terms of robots' trustworthiness~\cite{salem2015would,law2017wizard,roesler2020effect,hsu2021attitudes}, perceived intelligence~\cite{takayama2011expressing}, perceived likeability~\cite{mirnig2017err}, anthropomorphism~\cite{salem2015would}, reliability~\cite{salem2015would}, and ease of use~\cite{de2023we}. On the other hand, other research has suggested that AI mistakes could be considered as human-like characteristics, and therefore making the interactions more engaging~\cite{honig2018understanding,law2017wizard} and pleasurable~\cite{mirnig2017err,honig2018understanding,gompei2015robot}, mitigating the potential negative effect AI mistakes might have on users' perceptions of the AI~\cite{honig2018understanding,buhrke2021making,mirnig2017err}. 
Scholars conjectured that people's perception after encountering robot failures might be attributed to various factors such as types of failures, failure severity, or the nature of the task (e.g., formal workplace or social and personal)~\cite{sarkar2017effects,honig2018understanding,rossi2017timing}. Examining people's perception changes after encountering AI mistakes across contexts could better inform the design of AI's repair and recovery strategies. 

Following this line of research, others investigated AI's repair and mitigation strategies, focusing mostly on conversation breakdown during human-chatbot interactions~\cite{ashktorab2019resilient,mahmood2022owning} and human-robot interactions~\cite{honig2018understanding}.~\citeauthor{honig2018understanding} [2018] provided a comprehensive summary of four types of mitigation strategies during interaction breakdowns: (1) Setting expectations by forewarning users about potential flaws~\cite{lee2010gracefully,honig2018understanding,kocielnik2019will}, (2) Communicating errors properly by incorporating elements such as apologizing~\cite{mahmood2022owning}, humor~\cite{clausen2023exploring}, or through further explanations~\cite{ashktorab2019resilient}, (3) Asking users for help, and (4) Mix and match of different mitigation strategies listed above. One particular mitigation strategy that has gained popularity is explainable AI. However, empirical studies have suggested that common explanation strategies were not as effective as expected given that the explanations did not take into account people's domain expertise~\cite{wang2021explanations} and AI knowledge such as intuitions and beliefs~\cite{chen2023understanding}. This has sparked the area of human-centered explainable AI~\cite{miller2019explanation,liao2021human,ehsan2020human} that aims at presenting explanations relative to people's knowledge, capabilities, and beliefs.

%% file: 3studyoverview.tex
\section{Study Overview}
\begin{figure}[t]
    \centering
    \includegraphics[width=\textwidth]{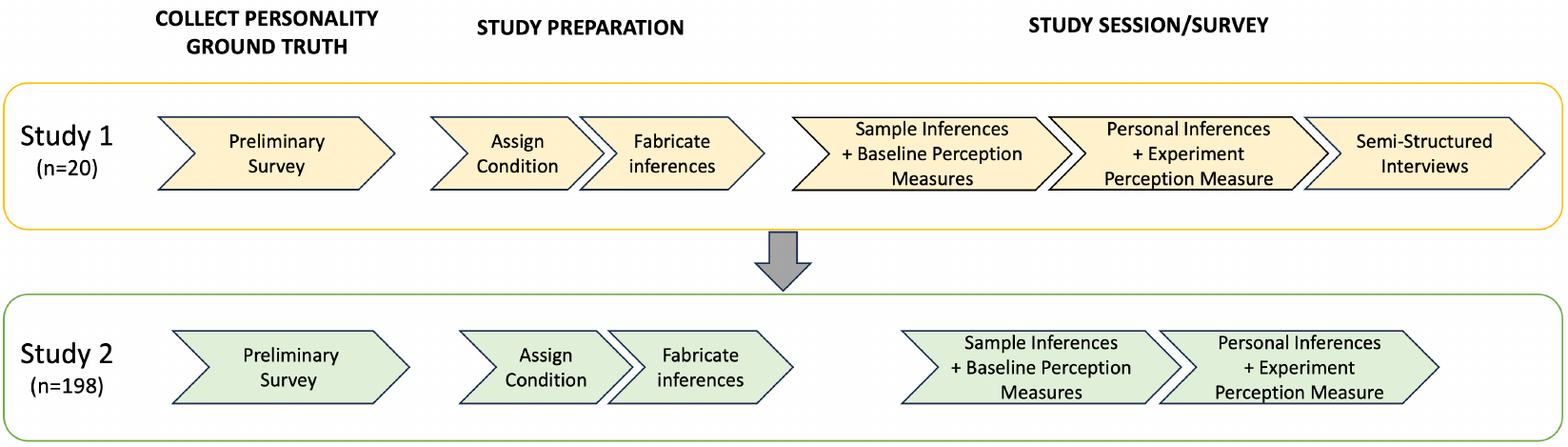}
    \caption{Study flow diagram that shows the procedures of Study 1 and Study 2. Study 2 occurred after Study 1 was concluded. All personal inferences shown to participants were either accurate or inaccurate based on the condition assigned to the participants.}
    \label{fig:study_procedure}
\end{figure}

To examine people's reactions and perceptions of AI after encountering AI misrepresentations, we conducted two studies using a mixed-methods approach. The first study (Study 1) focuses on qualitatively exploring students' perceptions and reactions to AI after encountering AI misrepresentations, and the second study (Study 2) focuses on quantitatively examining the factors, specifically AI literacy, contributed to the variations in students' perceptions of the AI after encountering AI misrepresentations. Given that AI output is often non-deterministic and thus difficult to control and manipulate, we took a Wizard-of-Oz approach to fabricate and control the accuracy of the AI inferences--- in both studies, human researchers fabricated all the AI inferences based on participants' personality ground truth, collected in the preliminary survey in each study. To prevent raising participants' suspicions when shown fabricated AI inferences, they were not presented with the inferences until at least a week after they completed the preliminary survey. 

To compare and contrast participants' reactions and perceptions of AI misrepresentations, we divided all participants into two conditions in both studies: participants in the accurate condition received accurate ``AI inferences,'' and the inaccurate condition received inaccurate ``AI inferences.'' In both studies, participants were asked to evaluate AI-generated inferences based on students' self-introduction paragraphs, which will be used by an AI agent named SAMI (stands for ``Social Agent Mediated Interaction'') to match them with potential teammates for a school project. 


In Study 1, we conducted user study sessions to understand students' reactions and perceptions (RQ1, RQ2) of the AI after encountering AI (mis)representations by showing them different SAMI inferences, including inferences about themselves. Based on our observations in Study 1, we revised some of the measurements and replicated Study 1 as a survey experiment without the semi-structured interview part. We conducted Study 2 on the Prolific crowdsourcing platform to obtain a larger sample to examine changes in students' perceptions (RQ2) and factors that contributed to those changes (RQ3). Both studies were approved by the IRB at the researchers' institute. Before Study 1, we conducted a pilot study with 14 people to test out the study procedure and the SAMI inference fabrication process. 

%% file: 4study1.tex
\section{Study 1: Understanding Students' Perceptions and Reactions to AI Misrepresentation}
In this section, we first describe the method we used in Study 1, including recruitment, study setup, and study procedure. We then talk about our data analysis process. Finally, we present our findings on students' perceptions and reactions to SAMI misrepresentations. 

\subsection{Study 1 Method}
\subsubsection{Study 1 Participant and Recruitment}
We conducted remote 60-minute user study sessions with 20 current students, all recruited from a large public U.S. technology institute. We recruited our participants by posting our recruitment message on the institute's Reddit and by broadcasting our recruitment email to different departments and programs, especially non-STEM programs, within the institute to increase sample diversity. Our recruitment message invited students to evaluate an AI agent SAMI that could perform team matching by drawing inferences from students’ self-introduction. Students signed up for the study by filling out a preliminary survey.

In the preliminary survey (see Supplementary Materials~\ref{supp:s1_prelim_survey}), students wrote a paragraph of self-introduction as a free-flowing essay to introduce themselves to SAMI. They then completed a 44-item Big Five personality survey measurement to provide ground truth of their personality. 
Students were then asked about level of technology proficiency (``Beginner'', ``Intermediate'', or ``Expert.''), general attitude towards AI technology on a scale of ``1-Very negative'' to ``5-Very positive'', as well as their demographic information, level of study, major, academic or professional background. 

We received 110 valid preliminary survey responses. We sent out follow-up invitations in batches to balance sample diversity and ended up with 20 participants in total. We then evenly assigned these participants to either the inaccurate or accurate condition (10 participants in each condition) while balancing the sample diversity in each condition. The median age of our participants was 20 years old. There were 15 undergraduate students and 5 master's students in our study. Their gender breakdown is as follows: women (n=10), men (n=7), non-binary (n=2), one participant did not report their gender. Our participants came from a variety of majors, with mostly intermediate tech proficiency, and varying general attitudes toward AI. Table~\ref{tab:s1_participant} in Appendix~\ref{appendix:s1participant} shows the information for all the participants in Study 1.

\subsubsection{Fabricating SAMI Inferences} \label{fabrication}


We took a Wizard of Oz approach and let human researchers fabricate SAMI's personality inferences to control for inference accuracy. The fabrication of SAMI inferences was based on (1) participants' ratings of each statement on the Big Five personality test, and (2) the condition (either accurate or inaccurate) the participant was assigned to. Participants in the accurate condition would receive accurate inferences, which would consist of all the statements they rated as ``4-agree'' or ``5-strongly agree'' in the personality test, or the reverse of the statements they rated as ``1-strongly disagree'' or ``2-disagree'' in the personality test; Participants in the inaccurate condition would receive inaccurate inferences, which would consist of all the statements they rated as ``1-strongly disagree'' or ``2-disagree'' in the personality test, or the reverse of statements that they rated as “4-agree” or “5-strongly agree” in the personality test. We illustrated this fabrication process in Fig.~\ref{fig:fabrication} in Appendix~\ref{appendix:fabrication}.

When selecting statements from the personality test, we picked the three dimensions with the most extreme scores, while excluding dimensions with a neutral score of three out of five. We also restricted the length of SAMI's inferences for each participant to about 10 statements to remove the potential effect of inference length on participants' perception of SAMI. In the Big Five personality test, some statements were inherently positive and some statements were inherently negative. To remove the potential effect of the sentiment of SAMI inferences on students' perceptions and reactions to SAMI, we composed each SAMI inference with about 40\% negative inference and 60\% positive inference. 

We composed an inference fabrication guideline to document these rules and procedures (see Supplementary Materials~\ref{supp:fabrication_rule}) and closely followed the guideline when fabricating SAMI inferences.

\subsubsection{Study 1 Remote User Study Session} 


We began each session by emphasizing to the participants that they should share their honest opinions, both positive and negative. 
Each study session had two parts. In the first part, participant was shown two samples of self-introductions and SAMI's inferences, then asked to fill out perception measurements (see Supplementary Materials~\ref{supp:s1_exp_survey}) to record their baseline perceptions of SAMI in terms of trust~\cite{gulati2019design} and social perceptions (likeability, perceived intelligence, anthropomorphism)~\cite{bartneck2009measurement}. Participant was then shown their own self-introduction and SAMI's inferences about them, then asked to fill out the same measurements. The goal was to record students' perceptions of SAMI immediately after each SAMI inference was presented to them. The two samples (see Supplementary Materials~\ref{supp:s1_protocol}) were written by real students from our pilot study. SAMI's inferences about the sample students were fabricated in the same way as described in section~\ref{fabrication}. SAMI's inferences for one of the samples were generated to be inaccurate, and the other one accurate. 
The samples were shown in random orders to remove potential order effects. 

The second part of the session was a semi-structured interview to go through participants' reactions and perceptions of SAMI after encountering AI (mis)representations. During the interviews, we asked participants to walk us through their reactions to each of the SAMI inference, what they thought about SAMI after seeing each inference, and how they thought SAMI extracted the inferences. Throughout the interview, we used the perception measurements as a probe to provide basis for them to elaborate on their perceptions of SAMI after seeing each inference. We then debriefed the participants about the fabrication process of SAMI's inferences. We addressed any comments and questions that participants had and then provided compensation of USD \$25 gift card. All the study sessions were conducted remotely on Zoom and all lasted around 60 minutes. The session and interview protocol is attached in Supplementary Materials~\ref{supp:s1_protocol}.

\subsection{Study 1 Data Analysis}
All sessions were video recorded and transcribed for data analysis. We adopted~\citeauthor{braun2006using}[2006]'s Reflexive Thematic Analysis (RTA) approach~\cite{braun2006using,braun2019reflecting} in our analysis. RTA encourages researchers to embrace their stance and assumptions during analysis. This differs from other qualitative data analysis approaches such as grounded theory and codebook approach that value objectivity and removing researcher bias, which does not allow enough flexibility for researchers to actively participate in the data analysis process.

Two researchers participated in the analysis to collaboratively discuss and iterate on themes that emerged from the data using RTA approach. We followed the analysis process outlined in~\citeauthor{braun2006using}[2006]. We generated our initial codes in two rounds by first dividing the 20 transcripts among the two researchers for independent coding, then swapped the transcripts for a second round of independent coding. After the initial codes were generated, the two researchers frequently met to discuss, review, and search for themes. The first round of analysis generated 16 domain categories (e.g., folk theories of SAMI) and 163 codes (e.g., believing SAMI's misrepresentation). By comparing our second round codes with the first round codes and domain categories, we distilled five themes (e.g., tendency to overtrust AI) and 27 codes (e.g., attributing human inference-making process to SAMI). We continued to review and search for larger themes and ended up distilling two bigger themes to understand students' perceptions and reactions to SAMI after encountering AI misrepresentations, which we present in detail in the next section.

\subsection{Study 1 Findings: Interpreting and Reacting to AI after Encountering AI (Mis)representation}

We first confirmed that our manipulation of SAMI inference accuracy was largely effective. We compared students' baseline accuracy rating of SAMI's inferences after seeing the samples and their experiment accuracy rating of SAMI's inferences after seeing their own inferences. We found that in the accurate condition, the accuracy rating largely increased from baseline rating to experiment rating, with a median increase of 1 out of 5; in the inaccurate group, the accuracy rating largely decreased from the baseline rating to the experiment rating, with a median decrease of 0.5 out of 5. However, it is worth mentioning that some participants' accuracy ratings did not change between baseline and experiment (n=1 in the accurate condition, n=3 in the inaccurate condition).

Next, we present our findings from Study 1 to understand students' reactions and perceptions of SAMI after encountering AI (mis)representations. We first describe three rationales that participants adopted through knowledge acquired from SAMI (mis)representations to interpret how SAMI worked: SAMI works like a machine, a human, and/or magic. We then describe participants' reactions of over-trusting, rationalizing, and forgiving of SAMI misrepresentations, highlighting that these reactions were highly connected to the rationales that we pinpointed. 



\subsubsection{Interpreting SAMI: Machine? Human? Or Magic?}
Based on our interviews with the participants, we found that participants acquired new knowledge from SAMI (mis)representations. Such newly acquired knowledge prompted participants to adopt different rationales to interpret how SAMI worked: SAMI works like a machine, human, and/or magic. These rationales could co-exist at any given time, yet are bounded by participants' existing AI knowledge, tech proficiency, and how much they could make sense of SAMI's specific inferences. 

\textbf{``SAMI works like a machine.''} The first rationale participants held was that SAMI followed the typical ``input-processing-output'' machine working mechanism. About a quarter of the participants in our study, half of whom self-rated as ``Expert'' in terms of their tech proficiency, could clearly explain how SAMI came up with specific inferences by specifying their perceived SAMI knowledge base, training, input processing, and inference generation process. 
P23 described her speculation of how SAMI processed the self-introduction input to generate personality inferences:~\textit{``From a personality standpoint, I'd imagine that it's sectioning out facts and being like, people who do this are commonly [like] that. And then kind of filtering through several things that point to this person being generally trusting. So in this sample, this person want to travel, want to visit the seven wonders of the world, maybe all of that means they are generally trusting.''}

\textbf{``SAMI works like a human.''} We also observed the ``SAMI works like a human'' rationale surfacing among some participants, all of whom self-reported their tech proficiency as ``Intermediate.'' These participants believed that SAMI drew personality inferences in similar ways as how humans would do it. For instance, P14 described how SAMI worked like humans when coming up with inferences:~\textit{``I think they are probably like us, where [certain things] kind of tick SAMI off. Like maybe people who love to travel can be a little bit more easily distracted, or maybe a little bit more careless, just because they have to be more go with the flow. ''} 
When P16 also described her rationale of how SAMI came up with inferences like humans would, she elaborated:~\textit{``So it's funny, because I know [humans and AI] are not the same at all. But I think like, who's making the AI technology--- people! So maybe the people try to make it think like how we think, even though it's not one and the same. Maybe whoever coded it tried to make it like human perception. ''}

~\textbf{``SAMI works like magic.''} When participants couldn't make sense of SAMI's unexpected inferences from the given self-introduction paragraph, they would sometimes resort to the belief that SAMI was a powerful, knowledgeable, ``big black box of magic.'' This rationale was mentioned by participants with varying levels of tech proficiency. Participants who adopted the magic rationale often could not clearly articulate how SAMI came up with the specific inferences, but instead loaded their answers with vague terms to emphasize the massive amount of training data SAMI might have access to or the ``magical power'' of AI to identify patterns that humans cannot see. For example, P36 who self-rated as an ``Expert'' in tech proficiency explained why he thought some of SAMI's inaccurate inferences about him were true:~\textit{``I was under the impression that it somehow drew inferences... since I didn't see where it got that from, I thought it was drawing inferences from different sources of information, and somehow using a pattern, like using a neural net or something. So I was like `Ok, sure maybe I can see that.' But if a person just told me that and I just don't see where they got it from, then I'd just think the person is being inaccurate. (P36)''}

\subsubsection{Reacting to AI Misrepresentation: Over-trusting, Rationalizing, and Forgiving}

After being shown SAMI's inaccurate personality inferences about them, participants displayed a range of reactions: some participants believed there was some truth to  SAMI's misrepresentation; some participants rationalized it and blamed themselves instead; some participants were forgiving of SAMI's mistakes. These reactions to SAMI misrepresentations seemed to be connected to the rationale participants adopted after acquiring new knowledge from SAMI misrepresentations about them. 

\textbf{Over-trusting SAMI Misrepresentations.} 
To our surprise, we found that most participants in the inaccurate condition often found some truth to SAMI's inaccurate inferences about them. Given that we made every SAMI inference to be intentionally inaccurate and the complete opposite of participants' personality ground truth, we expected most participants in the inaccurate condition to rate SAMI's inferences about them as ``1-Not accurate at all.'' However, only two out of 10 participants did so--- rest of the participants displayed varying levels of trust in SAMI's inaccurate inferences about them, ranging from two to ``3-Somewhat accurate''

These participants often fell into the trap of the Barnum effect~\cite{dickson1985barnum}, a cognitive bias phenomenon that people tend to believe in personality descriptions of them as customized to them, when in fact, these descriptions are often vague and general. When presented with SAMI's inferences about them, participants had a tendency to look for evidence in their daily lives to support SAMI's claims. 
In addition, believing in the authority of the evaluator could make the Barnum effect stronger~\cite{dickson1985barnum}. We found that participants in the inaccurate condition displayed a tendency to perceive SAMI as an authority figure that was ``smarter and more powerful'' than them, with opaque working mechanism. 

This echoed with the ``SAMI works like magic'' rationale that they don't know how AI works, but it just should work. This rationale was even present when participants were confident that SAMI's inferences were very inaccurate. For example, P28 rated SAMI's inferences as very inaccurate, yet still took a huge sigh of relief when we told her SAMI's inferences about here were fabricated to be inaccurate:~\textit{``I feel better now because I was worried. I was worried that, because I don't see myself from the outside so I was worried that SAMI was real and kind of accurate, and this is how people see me. ''} She further explained about her reasoning:
\textit{``I guess partly because it's an AI. Because AI is a lot smarter than me. And I don't consider myself to be a computer science expert. AI is so unfamiliar to me that I don't really know what is the line between true and false. I don't know when to trust it or not to. I think with SAMI, I was erring on the side of being not trusting but then deep inside, I was like, `this is an AI, it's really smart, it should be able to know.' I think that's why I felt like I could trust it. After I saw the two samples and then when it came to mine, I started to feel like something was wrong. Uhm, but then deep down I was like, `ok, well, maybe it is right.'''}

\textbf{Rationalizing SAMI Misrepresentations.} Besides participants' tendency to over-trust AI misrepresentations, we also found that participants had a tendency to rationalize AI mistakes, or ``find excuses'' for SAMI's inaccurate inferences. When participants noticed that SAMI's inferences might not be accurate for either the students in the sample or themselves, some participants would ``justify'' SAMI's inaccuracies, citing it could be a result of the nature and quality of the self-introductions. This echoed with the ``SAMI works like a machine'' rationale that participants believed ``unqualified'' input would hurt the machine's output. For instance, when P29 spotted SAMI's inaccurate inferences in one of the samples, she said,~\textit{``I feel like the person was fairly broad in what they wrote. And so that also could have influenced SAMI's response. Like it didn't necessarily go too in-depth, [...] It also went from learning languages, study abroad, and then the career of helping people. That could have thrown SAMI off a little bit as well, seeing as it like, it bounced around a little bit more. ''} P36 also said that SAMI might have made inaccurate inferences about him due to him not mentioning certain things in his self-introduction:~\textit{```You can be somewhat careless.' I would say that is like very inaccurate. I'm a very meticulous person. But I guess I never really touched on that anywhere in my response...''}

\textbf{Forgiving SAMI Misrepresentations.} Many participants were forgiving of SAMI's misrepresentations of either themselves or students in the samples. This echoed the ``SAMI works like a human'' rationale in that this reaction is analogous to how they would react to human mistakes. After encountering SAMI misrepresentations, participants attributed good intentions and efforts to SAMI. Following this line of reasoning, participants believed that just like human mistakes, SAMI's mistakes could be forgiven considering its good intentions and efforts. This reaction was presented in more than half of the participants in the inaccurate condition. After reviewing SAMI's misrepresentations of themselves, they believed that SAMI was ``trying'' and well-intentioned, but just not as capable as they expected. Some participants in the accurate condition, noticing that SAMI did not make perfect inferences in the samples, were still hopeful about SAMI despite its mistakes. P29 said,~\textit{``[I'm thinking of] a little kid where they want to help out and be like `oh my gosh, look what I did' and presented it to you and be like `Look, I'm trying to be helpful.' But it's not always the most helpful or accurate in this situation. I was thinking that [SAMI] is trying to help, it is generating those responses, but at the same time, it just might not be the most accurate.''}

%% file: 5study2.tex
\section{Study 2: Examining the Changes in Students' Perception of AI after Encountering AI Misrepresentations}
To explore the changes in students' perceptions of AI (RQ2 and RQ3) after encountering AI misrepresentations, we conducted a survey experiment on Prolific and solicited a larger sample to quantify the changes in student perceptions as well as contributing factors to the changes. Similar to Study 1, we first deployed a preliminary survey to collect information for fabricating students' personality inferences. Then after a week, we followed up with an experiment survey to show participants SAMI's inferences and measured their perception changes of SAMI. Based on our observations from Study 1, we added a general AI literacy measurement in the preliminary survey to examine AI literacy as a potential factor contributing to the changes in students' perceptions of AI after encountering AI misrepresentation. 


\subsection{Study 2 Study Design}

We deployed the preliminary survey (see Supplementary Materials~\ref{supp:s2_prelim_survey}) on Prolific and recruited current students above age 18 of all study levels in the United States. Participants were told that the goal of this study was to understand students' perception of a team-matching AI agent. In the preliminary survey, participants provided their self-introduction paragraph, responses to the personality test and an AI literacy questionnaire, as well as their demographic, background, and team project experience information. Participants who completed the preliminary survey were compensated with \$3 USD. The median completion time was 11 min and 41 seconds. 

250 Prolific participants filled out our preliminary survey on Qualtrics. We removed seven participants whose self-introductions were suspected to be generated by generative AI tools such as ChatGPT. Given our emphasis on AI misrepresentation in Study 2, we assigned about 70\% of the participants to the inaccurate condition and about 30\% of the participants to the accurate condition. We then wrote a Python script to fabricate SAMI's inferences for the 243 remaining participants following the same fabrication procedure and guidelines in Study 1. By following the 60\% positive and 40\% negative inference ratio rule, we removed 21 participants since their fabricated inferences were either too positive or too negative based on their personality ground truth. This resulted in 222 participants remaining (n=157 in inaccurate condition, n=65 in accurate condition), all of whom were invited to participate in the experiment survey. 

The experiment survey (see Supplementary Materials~\ref{supp:s2_exp_survey}) followed the Study 1 procedure by showing the same two samples and prompted participants to answer their perceived accuracy of SAMI inferences, as well as filling out measurements of their trust and social perceptions of SAMI. Participants then retrieved SAMI's inferences about them by entering their Prolific ID on a website that we created (see Appendix~\ref{appendix:s2website}), and then filled out the same perception measurements. Finally, a debriefing form informed participants about the real purpose of the study and how SAMI's inferences were fabricated. Participants were compensated with \$3 USD for completing the experiment survey, and an extra \$4 USD bonus for completing both surveys. We received 211 responses (n=151 for inaccurate condition, n=61 for accurate condition) for the experiment survey. The median completion time was 12 min and 43 seconds. A description of the perception measures and general AI literacy measures used in Study 2 can be found in Appendix~\ref{appendix:s2measures}.

After we concluded the data collection, we removed 13 participants' data due to extremely fast completion time (less than six minutes) and obvious contradictions in their responses. We ended up with 198 participants' data for the experiment survey, with 57 participants in the accurate condition and 141 participants in the inaccurate condition.

\subsection{Study 2 Participant Summary}

Our final participant pool (n=198) has an average age of 31.3±11.12, ranging between 18-74 years old. 42.9\% were women, 53.5\% were men. Most students were at the undergraduate level (n=146, 73.7\%). 46.5\% students in non-STEM major, 52\% in STEM major. Participants were relatively familiar with AI, with an average of 14.3±4.50 out of 25 on overall AI literacy.
Participants were generally experienced in team projects at school, having participated in an average of 12 team projects (median=5, SD=25.8, range=0--300). Participants held a relatively positive attitude about their team project experience, reflected in an average rating of 3.7 ± 0.88 out of a five-point Likert scale. 
We provided more details about the participants' demographic and personality in Appendix~\ref{appendix:s2_participant}.


\subsection{Study 2 Data Analysis}
We first calculated the changes in students' perceptions of SAMI (overall trust, anthropomorphism, perceived intelligence, likeability) by taking the difference between students' baseline perceptions ($B_p$) and experiment perceptions ($E_p$): $\Delta= E_p - B_p$. We then coded condition as 0 (accurate condition) and 1 (inaccurate condition) during analysis.

To understand the effect of AI misrepresentation on students' perception changes of SAMI (RQ2), we performed four sets of linear regressions with the $\Delta$ of each perception construct as the outcome and the participants' condition as the independent variable. We controlled for age, gender, study level, major (STEM or non-STEM), number of projects, project experience rating, and five personality dimensions. Equation~\ref{eq1} describes our linear regression models, where$\Delta~\mathcal{P}$ refers to changes in overall trust, anthropomorphism, intelligence, and likeability. 
\vspace{-0.1in}
\begin{multline}
\label{eq1}
\small
    \Delta\mathcal{P} \sim Condition + Age + Gender + Study Level + Major + Project Count + Project Experience \\+ Extroversion + Neuroticism + Agreeableness + Openness + Conscientiousness
\end{multline}
\vspace{-0.2in}

To understand the moderating effect of AI literacy (RQ3), we again performed four sets of linear regression models, but this time incorporating AI literacy. Specifically, we performed two versions of linear regression for each model: one that included AI literacy (Equation~\ref{eq2}), and another included AI literacy plus an interaction effect between AI literacy and condition (Equation~\ref{eq3}). We included the only significant covariate from prior models--- the Openness personality dimension. The outcome $\Delta \mathcal{P}$ is the changes in overall trust, anthropomorphism, intelligence, and likeability. 
\begin{equation}
    \label{eq2}
    \small
    \Delta\mathcal{P} \sim Condition + AI Literacy + Openness
\end{equation}
\vspace{-0.2in}

\vspace{-0.2in}
\begin{equation}
\label{eq3}
\small
    \Delta\mathcal{P} \sim Condition + AI Literacy + Condition * AI Literacy + Openness
\end{equation}
\vspace{-0.15in}

\subsection{Study 2 Findings}
Our results show that encountering AI misrepresentations had a significant effect on changes in students' overall trust, perceived intelligence, anthropomorphism, and likeability of SAMI. We also found that AI literacy could moderate the effect of AI misrepresentations on changes in students' overall trust in SAMI, but not their social perceptions of SAMI.

\subsubsection{The Effect of AI Misrepresentation on Students' Perceptions of AI} 
We first looked at students' perception changes before and after encountering AI misrepresentations. We found that students' baseline perceptions in both conditions were similar in terms of overall trust, intelligence, anthropomorphism, and likeability, with a difference in median of less than 0.4 between conditions (details in Appendix~\ref{appendix:s2effectonchanges}). We then plotted out the changes in each perception outcome in density plots (Fig.~\ref{fig:perception_changes_density} in Appendix~\ref{appendix:s2effectonchanges}) and noticed that many participants in the inaccurate condition decreased their perceptions after encountering AI misrepresentations compared to participants in the accurate condition. Therefore, encountering AI misrepresentations seemed to have a negative effect on students' perceptions of SAMI.

To further examine the effect of encountering AI misrepresentation on students' perceptions of SAMI, we built four linear regression models with changes in each perception as the outcome. Our models show that after controlling for demographics, team project numbers and experiences, as well as the personality dimensions, encountering AI misrepresentation had a significant effect on changes in students' perceptions of SAMI in terms of overall trust, perceived intelligence, anthropomorphism, and likeability (Table~\ref{tab:lm_perceptionchanges} in Appendix~\ref{appendix:s2effectonchanges}). Specifically, participants who encountered AI misrepresentations reported significantly lower overall trust in SAMI (Est.=-1.14, p < 0.001) in comparison to those in the accurate condition. We also found that participants who encountered AI misrepresentations reported significantly lower perceived intelligence (Est.=-0.92, p < 0.001), anthropomorphism (Est.=-0.50, p < 0.001), and likeability (Est.=-0.58, p < 0.001) compared to the participants in the accurate condition. This suggested that after people encountered AI misrepresentation, they were more likely to view the AI as less trustful, less intelligent, less humanlike, and less likable. 

\subsubsection{AI Literacy as a Moderator on the Effect of AI Misrepresentation on Students' Perceptions of AI} We then looked at the possible moderating effect of AI literacy on the effect of AI misrepresentation on the changes in students' perceptions of AI. Our base models that include an effect for AI literacy (Equation~\ref{eq2}) (Models 1a., 2a., 3a., 4a., in Table~\ref{tab:moderation} in Appendix~\ref{appendix:s2_literacy}) show that AI literacy alone does not have a direct relationship with the outcomes; however, our models that include an interaction effect between AI literacy and condition (Equation~\ref{eq3}) (Models 1b., 2b., 3b., 4b., in Table~\ref{tab:moderation} in Appendix~\ref{appendix:s2_literacy}) show that there is a significant interaction between AI literacy and condition on changes in overall trust (Est. = -0.06, p < 0.05). However, this interaction is not significant on changes in intelligence, anthropomorphism, and likeability. Detailed model results can be found in Table~\ref{tab:moderation} in Appendix~\ref{appendix:s2_literacy}. This suggests that students' AI literacy could have an effect on students' overall trust of AI after encountering AI misrepresentations; however, students' AI literacy does not have an effect on students' perceived intelligence, anthropomorphism, likeability of the AI after encountering AI misrepresentation. 

\begin{figure}
    \centering
        \begin{subfigure}[b]{0.24\textwidth}
            \centering
            \includegraphics[width=\textwidth]{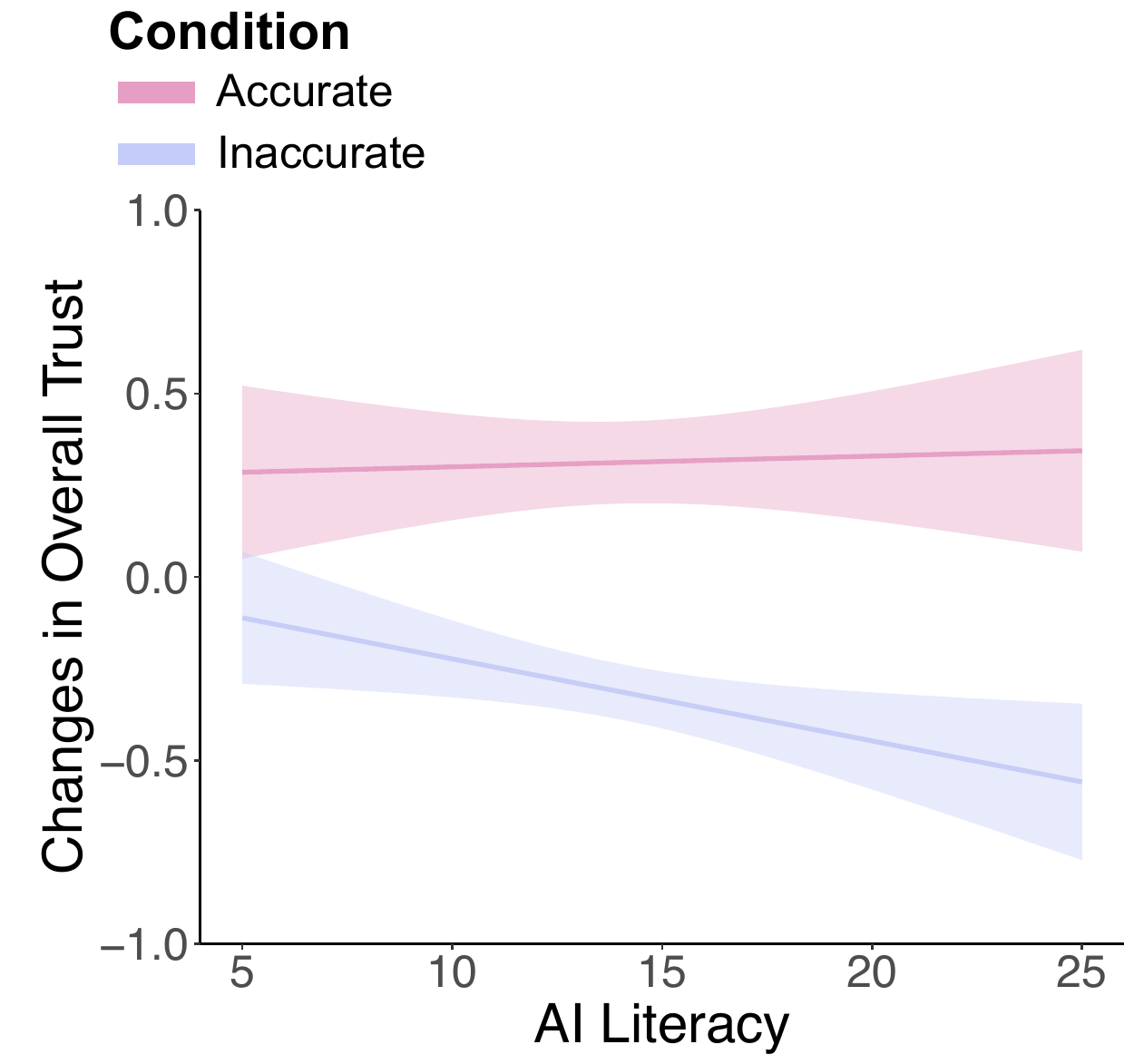}
            \caption{Overall Trust}
            \label{fig:interaction_trust}
        \end{subfigure}
        \hfill
        \begin{subfigure}[b]{0.24\textwidth}
            \centering
            \includegraphics[width=\textwidth]{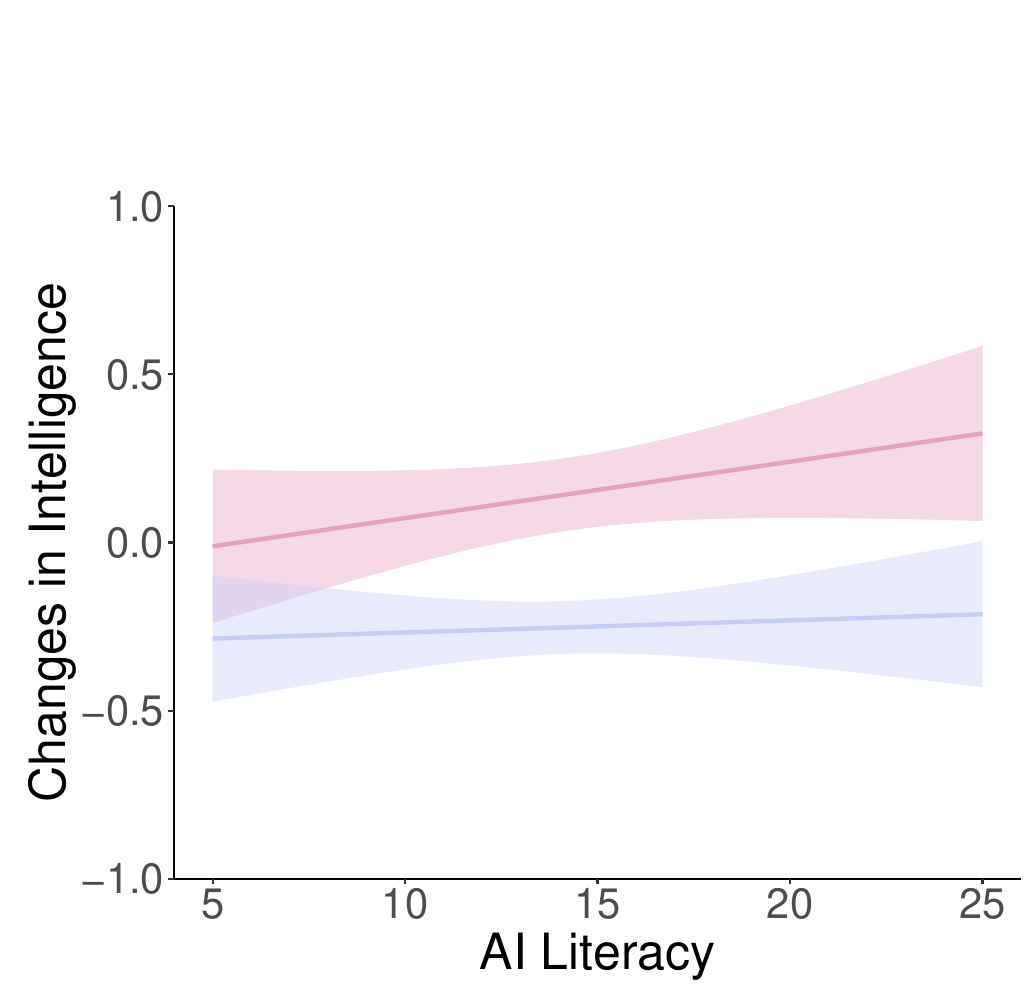}
            \caption{Intelligence}
            \label{fig:interaction_int}
        \end{subfigure}
        \hfill
        \begin{subfigure}[b]{0.24\textwidth}
            \centering
            \includegraphics[width=\textwidth]{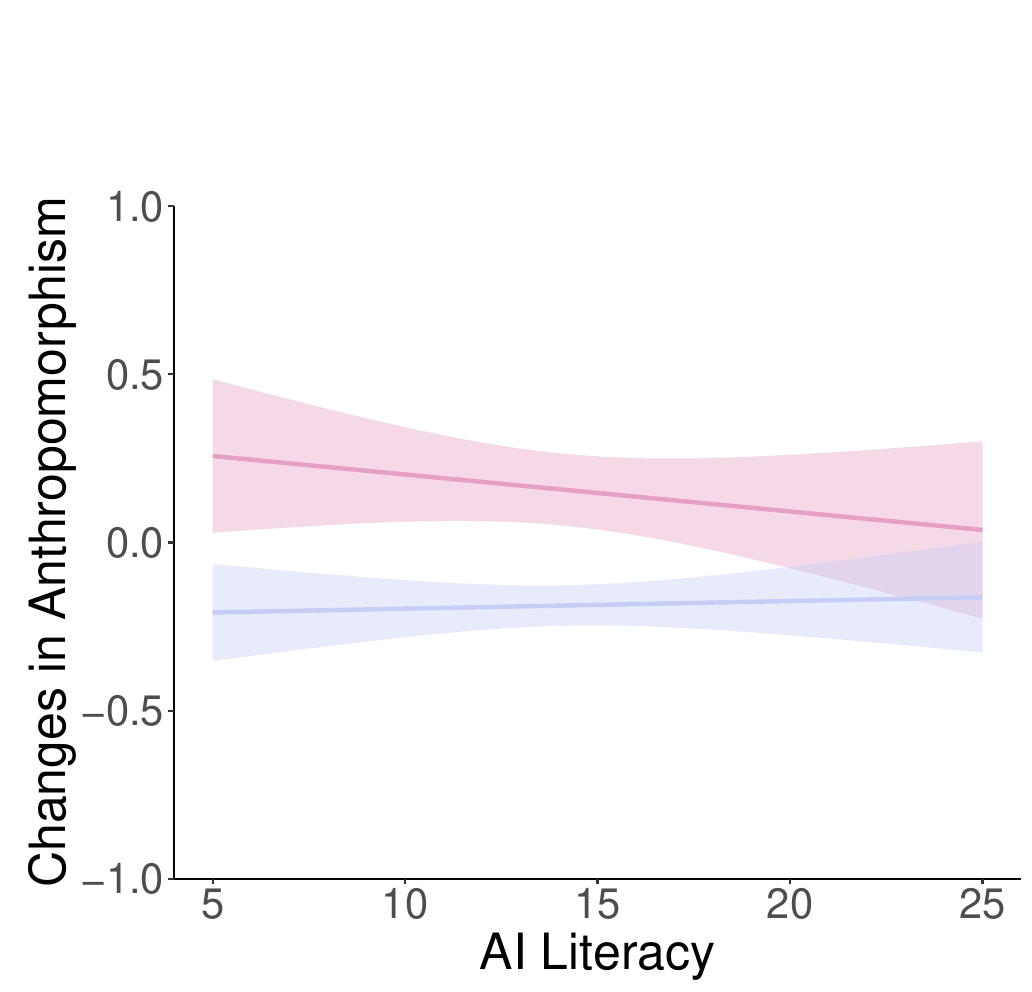}
            \caption{Anthropomorphism}
            \label{fig:interaction_anth}
        \end{subfigure}
        \hfill
        \begin{subfigure}[b]{0.24\textwidth}
            \centering
            \includegraphics[width=\textwidth]{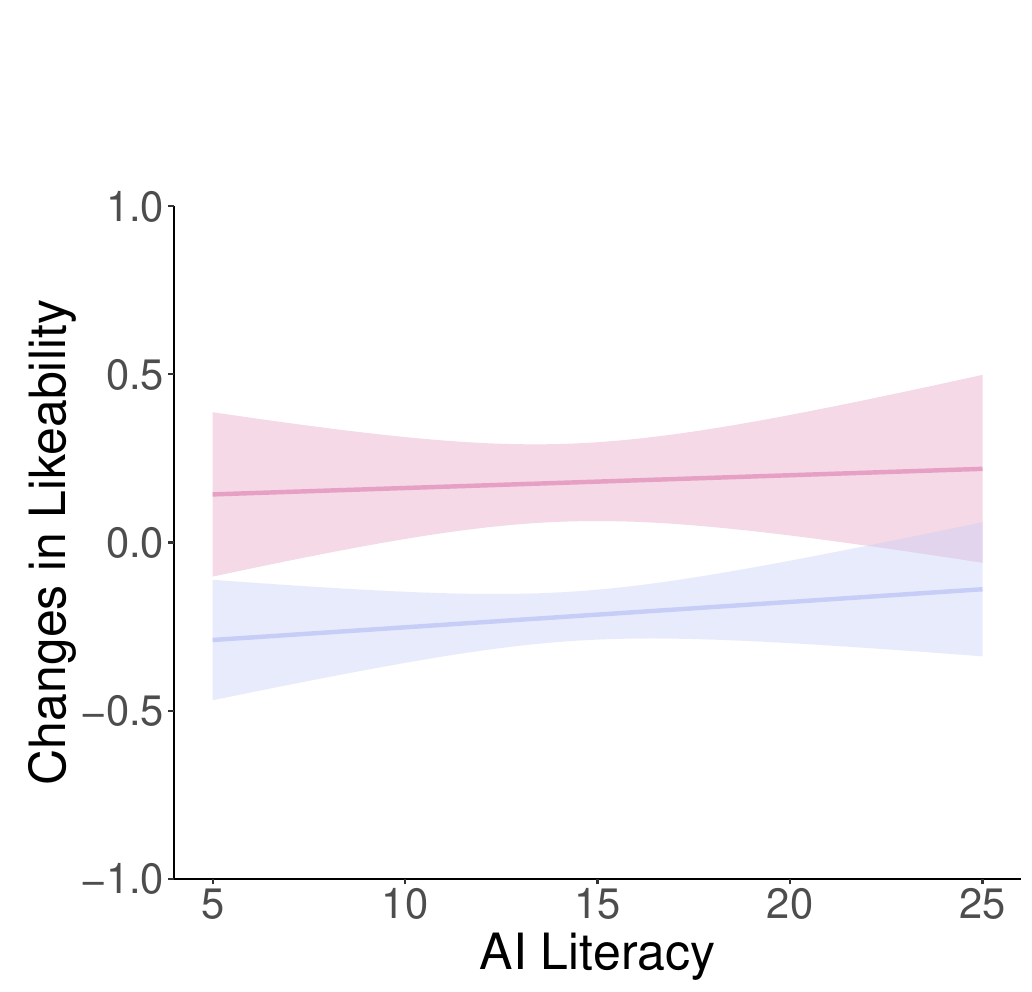}
            \caption{Likeability}
            \label{fig:interaction_like}
        \end{subfigure}
    \caption{(a) AI literacy significantly moderated the effect of AI misrepresentations on students' changes in overall trust of SAMI. (b) (c) (d) show that AI literacy does not significantly moderate the effect of AI misrepresentations on students' changes in perceived intelligence, anthropomorphism, and likeability of SAMI.}
    \label{fig:literacy_interaction}
\end{figure}

Fig.~\ref{fig:literacy_interaction} shows the interaction effect between AI literacy and condition on students' changes in perceptions. 
In Fig.~\ref{fig:interaction_trust}, students with higher AI literacy were more likely to change their overall trust in SAMI after encountering AI misrepresentations; students with lower AI literacy were less likely to change their overall trust in SAMI after encountering AI misrepresentations. Figures~\ref{fig:interaction_int},~\ref{fig:interaction_anth},~\ref{fig:interaction_like} show the non-significant interaction effect between AI literacy and condition, suggesting that students' levels of AI literacy has no significant effect on their changes in perceptions of intelligence, anthropomorphism, and likeability of the AI after encountering AI misrepresentations.

We then conducted post-hoc analysis to explore which specific dimensions of general AI literacy play an effect on students' changes in their overall trust in AI after encountering AI mistakes. We set up five linear regression models with students' changes in overall trust in SAMI as the outcome variable and included an effect of condition, each of the five AI literacy dimensions from the general AI literacy scale, and an interaction effect between condition and each AI literacy dimension. 
We controlled for the Openness personality dimension in all five models. We plotted out the interaction effect of each model (Fig.~\ref{fig:post-hoc_analysis} in Appendix~\ref{appendix:s2posthoc}) and we found a significant interaction effect between condition and students'~\textit{AI steps knowledge} (Est. = -0.23, S.E = 0.101, p < 0.05) on changes in students' overall trust in SAMI after encountering AI misrepresentations. This suggests that students with more general knowledge about AI's input, processing, and output are more likely to change their overall trust in AI after encountering AI misrepresentations. We also found a significant interaction effect between condition and students'~\textit{human actors in AI knowledge} (Est.=-0.22, S.E=-.111, p<0.05) on changes in students' overall trust in AI after encountering AI misrepresentations. Our result suggests that students with more knowledge of human involvement in the design and development of AI are more likely to change their overall trust in AI after encountering AI misrepresentations. Finally, we found a significant interaction effect between condition and students'~\textit{AI usage experience} (Est.=-0.32, S.E=0.112, p <0.01) and a significant effect of students'~\textit{AI usage experience} (Est.=0.19, S.E=0.086, p < 0.05) on changes in students overall trust in SAMI after encountering AI misrepresentations. This shows that students with more experience interacting and using different AI in their daily lives are more likely to change their overall trust in AI after encountering AI misrepresentations.

%% file: 6discussion.tex
\section{Discussion}
Through our mixed-methods approach, we offered insights on how people's existing and newly acquired knowledge of AI are highly connected to people's reactions and perceptions of AI after encountering AI misrepresentations. Specifically, we identified three rationales that participants adopted to interpret AI (mis)representations, reflecting participants' knowledge acquired from viewing AI outputs: AI works like a machine, a human, and/or magic. We found these rationales to be highly connected to participants' reactions of over-trusting, rationalizing, and forgiving of AI misrepresentations. Building on top of prior work that has suggested people's tendency of over-trusting and viewing AI as an authority~\cite{kapania2022because,warshaw2015can,hollis2018being}, we highlighted that these reactions and perceptions still persisted, and even exacerbated, when people encountered AI misrepresentations. We also empirically established that people's existing AI knowledge, i.e., AI literacy, can significantly moderate changes in people's trust of the AI after encountering AI misrepresentations, highlighting the importance of taking into account of people's knowledge and characteristics when building trustworthy AI systems~\cite{chen2023machine,chen2023understanding,ehsan2020human,schoeffer2022there}. In this section, we further unpack how people navigate AI fallibilities through their evolving AI knowledge. We then discuss implications for designing responsible mitigation and repair strategies by considering people's AI knowledge when AI fails. 





\subsection{Navigating AI Fallabilities Through Evolving AI Knowledge}
Our findings suggest that when facing AI mistakes such as AI misrepresentations, people acquire new knowledge about the AI, which could be consistent or inconsistent with their existing AI knowledge, prompting them to adopt a dominant rationale to explain such AI behaviors. Contrary to prior work that uses fixed heuristics to explain people's perceptions and reactions to AI systems~\cite{sundar2019machine,sundar2020rise,nass1994computers}, our findings suggest that people are constantly re-framing their perceptions
as they were presented with or discovered new information about the AI~\cite{devito2017platforms}. In Study 1, participants rarely stuck to one rationale about SAMI throughout, but kept navigating between different rationales, sometimes within a matter of seconds, as new information occurred to them. This suggests the instinctive nature of people's rationales~\cite{chen2023machine,chen2023understanding}--- we noticed that people rarely paused and reflected on their mental models~\cite{gero2020mental,wang2021towards} or folk theories~\cite{devito2017platforms,eslami2016first} of the AI before reacting to the AI output. This process of adopting various rationales, or concepts, to reconcile newly acquired knowledge with existing knowledge is similar to what was described as ``conceptual change''~\cite{ozdemir2007overview,chi1994things} in learning sciences, suggesting that people's re-framing of their perceptions
could be viewed as an evolving learning process about AI~\cite{wang2021towards}.

We note that people's rationales are often bounded by their existing AI knowledge, i.e., AI literacy~\cite{long2020ai,druga2017hey}. In Study 1, we observed that participants who self-reported as ``Expert'' in tech proficiency were more likely to adopt the machine rationale, whereas participants who self-reported as ``Intermediate'' were more likely to adopt the human rationale. Some participants also constantly referred back to their lack of AI literacy when interpreting SAMI's outputs, prompting them to adopt the magic rationale. Our findings also suggest that people's existing AI knowledge, like people's evolving rationales, are also subject to frequent changes. Our posthoc analysis showed that while dimensions such as AI steps knowledge and human actors in AI require formal or informal intentional learning about AI, people's AI usage experience is also a crucial dimension of AI literacy that could moderate changes in people's trust after encountering AI misrepresentation. This is also reflected in Study 1 where many participants mentioned their experience with ChatGPT when talking about their over-trust in SAMI and their magic rationale. This further emphasizes that people are constantly learning from their daily interactions with AI systems, which contributes to their evolving AI knowledge.

\subsection{Designing Responsible Mitigation by Considering People's AI Knowledge} 

While much of prior work has shown people's tendency to view AI as an authority~\cite{kapania2022because,warshaw2015can} and to overtrust AI-generated responses~\cite{warshaw2015can,gou2014knowme,hollis2018being}, our work, focusing on the scenario where AI makes mistakes, demonstrated that people would still over-trust, forgive, and even rationalize AI's obvious misjudgment in their most personal characteristics. We consider these responses to be highly connected to the rationale people adopted at the time, suggesting that some aspects of certain rationales could be harmful and even dangerous when generalized. This was illustrated by P28's reactions when she clearly recognized that SAMI's personality inferences about her were inaccurate, yet her perceived lack of AI knowledge prompted her to adopt the ``AI works like magic'' rationale and believed SAMI's misrepresentations could be true. This caused confusion, mental discomfort, and even self-doubt for P28. We therefore urge designers and developers of AI systems and algorithms that could infer people's personal traits such as personalities, emotions, preferences, to be aware and cautious of the potential harm to the user when such AI systems exhibit ``gaslighting'' behaviors, and design appropriate mitigation strategies to mitigate the potential harms. Based on our findings, we propose to incorporate people's evolving AI knowledge when designing mitigation strategies when AI fails. 

Existing repair and mitigation strategies mostly focus on adding social elements such as politeness, apologizing, or setting user expectations early in the interactions~\cite{honig2018understanding,lee2010gracefully,mahmood2022owning}. These strategies, without considering users' evolving AI knowledge, could risk eliciting undesired outcomes such as reinforcing people's social behaviors towards AI~\cite{nass1994computers} or confusing the users about AI's true capabilities as they learn. Echoing with prior work that people's beliefs and intuitions should be taken into consideration when designing AI explanations~\cite{miller2019explanation,liao2021human,ehsan2020human,long2020ai} and repair strategies~\cite{ashktorab2019resilient,chen2023understanding}, we provided a specific set of rationales and encourage future work to explore techniques that could allow automatic identifications of users' rationales in real-time. One mitigation strategy could be to provide explanations tailored to the specific rationale that people adopted at the time. For instance, if a user adopted the magic rationale, the AI could provide explanations to nudge the user to adopt the machine rationale to reduce overreliance. 

Additionally, people's existing AI knowledge, i.e., AI literacy~\cite{long2020ai,druga2017hey}, could be leveraged to approximate the ``cost'' of an AI misrepresentation when providing customized mitigation strategies. Our study found that people with lower AI literacy were not affected much by their trust in AI after encountering AI misrepresentation. Given that they still trust the AI after the AI erred, they might not provide any feedback, which could give the AI developers a false sense that the AI was working fine, leading to long-term repercussions. By contrast, while more AI literacy in a user can seem favorable, our study showed that it could lead to more extreme changes in trust perceptions in AI after AI misrepresentations. People with more AI literacy might abandon the system after encountering AI failures, which makes recovery non-trivial. By considering AI literacy as a factor, we have an opportunity to consider different repair strategies by estimating the different effects of AI misrepresentations for people with varying AI literacy levels.

%% file: 7conclusion.tex
\section{Limitations}
Our work has some limitations. First, we only studied people's perception changes after encountering a one-time AI misrepresentation in a very short period of time. People's perceptions and rationale change behaviors could be more stable in the long run~\cite{wang2021towards}, and we encourage future research to replicate our studies in longer-term settings and contexts. Second, while we incorporated many relevant covariates in our models, other factors that were not modeled could impact people's perceptions of AI, e.g., people's attitudes towards specific types of AI. Third, all study participants in study 1 were recruited from a large public technical institute and, despite our efforts to target recruit participants from non-STEM disciplines, participants in study 1 might have more knowledge and exposure to AI technologies than average college students. All of our study participants were recruited from the U.S. and may only represent Western attitudes and perceptions of AI~\cite{kapania2022because}. Future studies should consider replicating our study in non-western regions to understand the cultural differences in changes in perception and reactions of AI after encountering AI mistakes.

\section{Conclusion}
We reported on semi-structured interviews (n=20) and a large survey experiment (n=198) with college students in the United States to understand their perceptions and reactions of AI after encountering personality misrepresentations in AI-facilitated team matching. Through the two studies, we highlighted the critical role of people's existing and newly acquired knowledge of AI in shaping people's reactions and perceptions of AI after encountering AI misrepresentations. We emphasized the ever-evolving nature of people's AI knowledge acquired from viewing AI outputs by pinpointing three rationales that people adopted to interpret AI (mis)representations: AI works like a machine, a human, and/or magic. These rationales are bounded by people's existing AI knowledge and are highly connected to people's tendency to over-trust, rationalize, forgive AI misrepresentations. We also found that people's existing AI knowledge, i.e., AI literacy, could significantly moderate changes in people's trust in AI after encountering AI misrepresentations. We discussed how people navigate AI fallibility through their evolving AI knowledge and provided implications for designing and developing responsible mitigation strategies that consider people's evolving AI knowledge to reduce potential harms when AI fails.

%% file: 8appendix.tex
\clearpage
\renewcommand{\thefigure}{\thesection\arabic{figure}}
\renewcommand{\thetable}{\thesection\arabic{table}}
\setcounter{figure}{0}    
\setcounter{table}{0}   

\section{SAMI Inference Fabrication Process} \label{appendix:fabrication}
\begin{figure}[h]
    \centering
    \includegraphics[width=\textwidth]{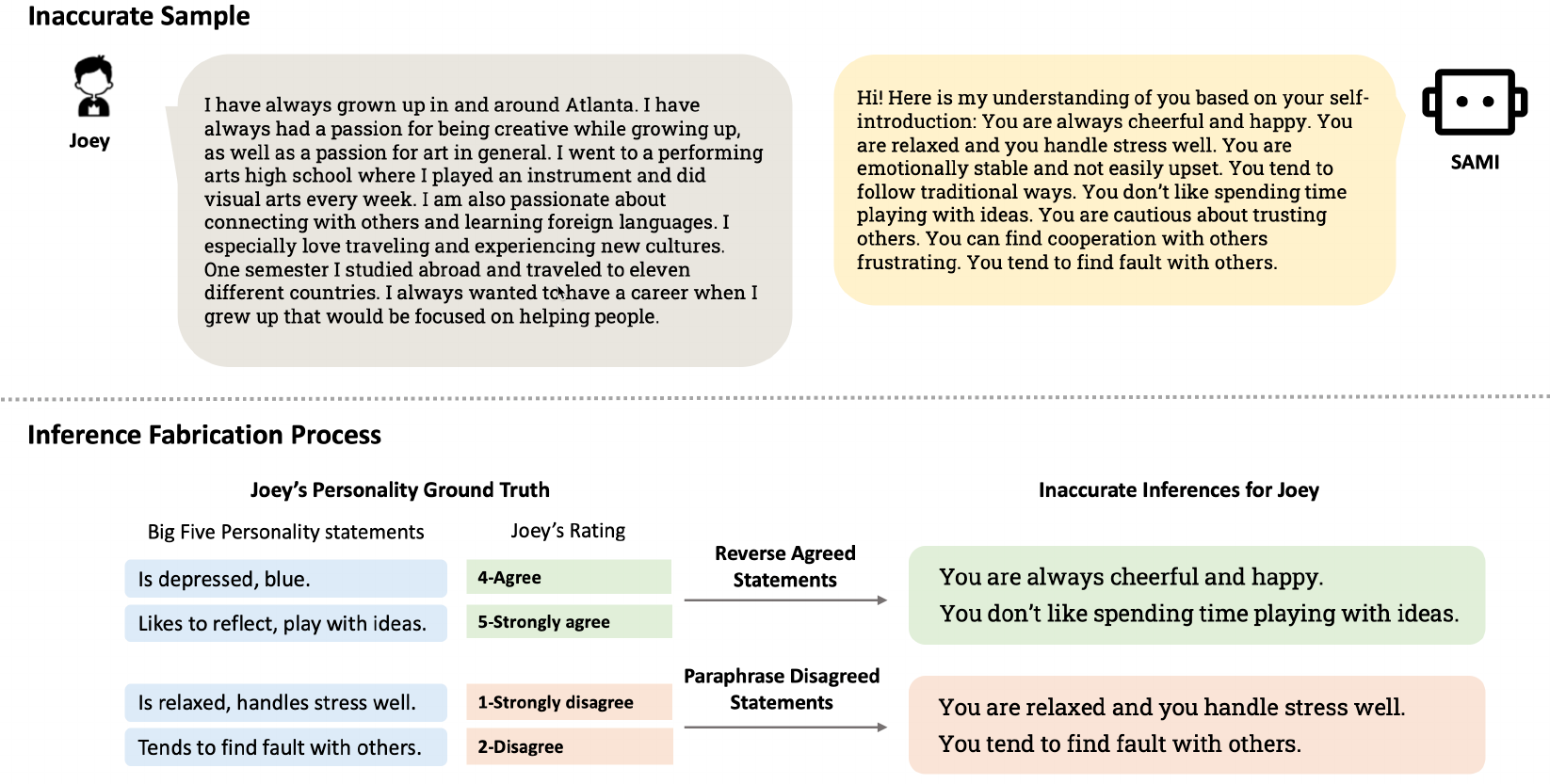}
    \caption{This figure shows the sample and our inference fabrication process for the sample student. The top half of this figure shows one of the samples we showed to the participants that is inaccurate. The bottom half of this figure shows how we utilized participants' personality ground truth filled out in the preliminary survey to fabricate inferences for them based on the condition they were assigned.}
    \label{fig:fabrication}
\end{figure}

\clearpage
\section{Study 1: Participant Information Table} \label{appendix:s1participant}
\begin{table}[h]
    \centering
    \small
    \caption{Study 1 participant information. In the Gender column, ``W'' stands for ``Woman'', ``M'' stands for ``Man'', ``NB'' stands for ``Non-Binary.'' In the Level of Study column, ``UG'' stands for ``Undergraduate.'' In the Major column, ``Eng.'' stands for ``Engineering'', ``Comp.'' stands for ``Computational.'' In the Tech Proficiency column, participants self-reported their technology proficiency as ``Beginner'', ``Intermediate'' or ``Expert.'' In the Attitudes Toward AI column, participants self-reported their attitudes toward AI on a scale of ``1-Very Negative'' to ``5-Very Positive.''}
    \begin{tabular}{clllllll} \toprule
     \textbf{Condition} & \textbf{ID} & \textbf{Age}  &   \textbf{Gender}  &   \pbox{1cm}{\textbf{Study \newline Level}}  &   \textbf{Major} & \pbox{2cm}{\textbf{Tech \newline Proficiency}}     &       \pbox{3cm}{\textbf{Attitudes \newline Toward AI}}\\ \midrule
     \multirow{10}{5em}{\centering Accurate Condition (n=10)} &   P14    &  22     &    W    &    UG    &    Psychology     &       Intermediate    &   Neutral to Positive \\
               & P19  &  18    &   W   &    UG    &    Psychology    &    Intermediate      &       Neutral to Negative \\
               & P22  &  28    &   W   &    Master    &    Digital Media    &    Intermediate       &       Neutral \\ 
               & P26  & 21      & M     & Master    & Digital Media     & Intermediate      &   Neutral \\
               & P29    & 19    & W     &   UG      & Neuroscience      & Intermediate  & Neutral to Negative \\
               & P30    & 31    & W     &   Master     & HCI      & Expert  & Neutral to Positive \\
               & P33    & 19    & M     &   UG     & Chemical Eng.      & Intermediate  & Neutral to Positive \\
               & P34    & 19    & NB     &   UG     & Comp. Media      & Intermediate  & Very Positive \\
               & P37    & 20    & M     &   UG     & CS      & Expert  & Very Positive \\
               & P40    & 19    & M     &   UG     & Computer Eng.     & Intermediate  & Very Positive \\   \midrule
    \multirow{10}{5em}{\centering Inaccurate Condition (n=10)} & P16    &   19  & W     &   UG  &   Psychology  &   Intermediate    &   Neutral to Positive \\
    & P17  &  21    &   W   &    UG    &    Psychology    &    Intermediate      &       Neutral \\
    & P21  &  19    &   NB   &    UG    &    Psychology    &    Expert      &       Neutral to Positive \\
    & P23  &  20    &   W   &    UG    &    Psychology    &    Intermediate      &       Neutral to Negative \\
    & P24  &  23    &   W   &    Master    &    HCI    &    Intermediate      &       Neutral to Positive \\
    & P28  &  20    &   W   &    UG    &    Business Admin    &    Intermediate      &       Neutral to Positive \\
    & P35  &  22    &   M   &    Master    &    CS    &    Expert      &       Neutral to Positive \\
    & P36  &  22    &   M   &    UG    &    Industrial Eng.    &    Expert      &       Neutral \\
    & P39  &  21    &   No report   &    UG    &    Biology    &    Intermediate      &       Neutral to Positive \\
    & P41  &  18    &   M   &    UG    &    Computer Eng.    &    Expert      &       Neutral to Positive \\
    \bottomrule
       
    \end{tabular}
    
    \label{tab:s1_participant}
\end{table}

\section{Study 2: Survey Measures} \label{appendix:s2measures}
In this section, we describe the three scales that we used in the preliminary survey and the experiment survey: the General AI Literacy Scale measures AI literacy as our moderator, the Human-Computer Trust scale measures students' trust in SAMI as one of the four outcomes, and the Godspeed scale for human-robot interaction measures perceived intelligence, anthropomorphism, and likeability of SAMI as the remaining three outcomes.

\textbf{General AI Literacy Scale: } This 13-item scale was developed and validated by~\citeauthor{pinski2023ai} to measure general AI literacy, which they interpreted as ``humans' socio-technical competences regarding AI''~\cite{pinski2023ai}. This scale consists of five dimensions: AI technology knowledge, human actions in AI knowledge, AI steps knowledge, AI usage experience, and AI design experience. The responses were recorded on a five-point Likert scale. Our correlation test suggested that the five dimensions were highly inter-correlated in our sample, with correlation ranging from 0.31 to 0.73. To avoid inflating our regression model, we decided to use an overall literacy score which is the sum of the score for each dimension. 

This general AI literacy scale measures five dimensions of general AI literacy:~\textit{AI technology knowledge}, which measures participants' knowledge regarding what makes AI distinct from other technology and the role of AI in human-AI interaction;~\textit{Human actions in AI knowledge}, which measures participants' knowledge of the role of human actors in human-AI interaction; ~\textit{AI steps knowledge}, which measures participants' knowledge about AI's input, processing, and output and each step's impact on humans;~\textit{AI usage experience}, which measures participants' use experience with AI;~\textit{``AI design experience''}, which measures participants' experience in designing and developing AI models and/or AI-driven products. 

\textbf{Human-Computer Trust Scale:} We measured students' trust in SAMI by using the ``Human-Computer Trust Scale'' developed and validated by~\citeauthor{gulati2019design}. This scale consists of 12 statements that can be customized to the specific AI technology being studied, and responses were recorded on a five-point Likert scale. This scale measures four dimensions of trust: perceived risk, benevolence, competence, and overall trust. Given that the scores for each of the dimensions were highly inter-correlated, with correlation magnitude ranging from 0.45 to 0.75, we took the score for overall trust to represent participants' trust in SAMI in our analysis. 

\textbf{Godspeed Scale for Human-Robot Interaction:} We measured students' social perceptions of SAMI by using the Godspeed scale developed and empirically validated by~\citeauthor{bartneck2009measurement}. This scale has been commonly used to measure users' social perception of AI agents in prior literature ~\cite{wang2021towards,mirnig2017err,honig2018understanding,salem2015would}. This scale is a semantic differential scale that asks participants to indicate their position on a scale between two bipolar words (e.g., Fake 1 2 3 4 5 Natural). We adapted the scale and took the three dimension measurements that were applicable to SAMI:~\textit{Anthropomorphism, Perceived Intelligence, and Likeability.}~\textit{Anthropomorphism} measures participants' level of attribution of human forms or human characteristics to the agent;~\textit{Perceived Intelligence} measures participants' perception of how smart or intelligent the agent is;~\textit{Likeability} measures participants' level of positive impression of the agent. 

\section{Study 2: Screenshot of the Website for Retrieving SAMI Inference} \label{appendix:s2website}
\begin{figure}[h]
    \centering
    \includegraphics[width=\columnwidth]{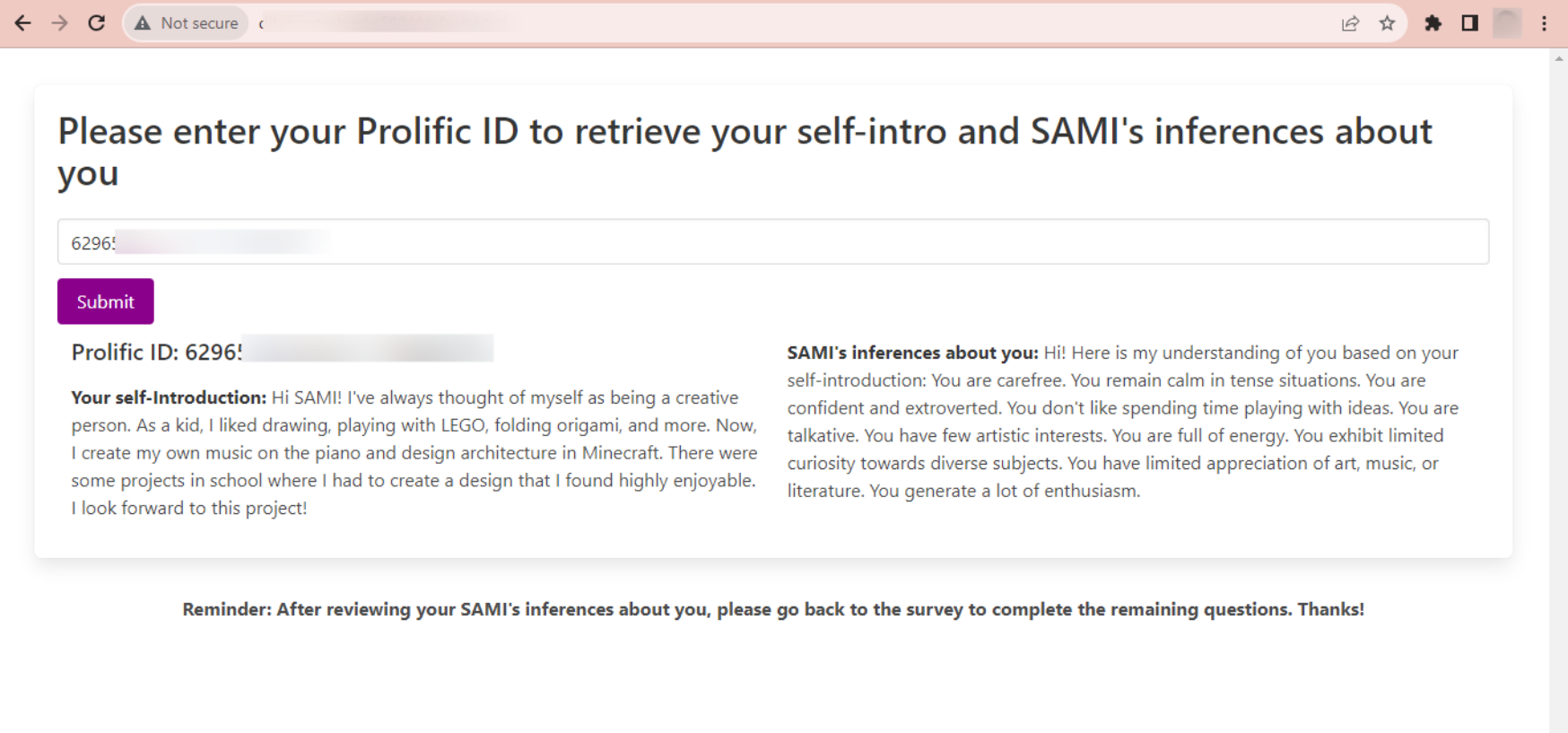}
    \caption{This is a screenshot of the website that we built for participants in Study 2 to retrieve SAMI's inferences about them by entering their Prolific ID.}
    \label{fig:website}
\end{figure}

\section{Study 2: Participant Information} \label{appendix:s2_participant}
In our final participant pool (n=198), the average age was 31.3 (median=28, SD=11.12, range=18-74). Among the participants, there were 42.9\% women (n=85), 53.5\% men (n=106), 0.03\% non-binary (n=5), 0.01\% prefer not to say (n=1), and 0.01\% prefer to self-describe (n=1, self-described as ``agender''). There were 73.7\% currently studying at the undergraduate level (n=146), 17.2\% at the master level (n=34), 0.02\% at the doctorate level (n=3), and 0.03\% described as other levels (E.g., Gen.Ed., associate degree). There were 46.5\% studying non-STEM major (n=92), 52\% studying STEM major (n=103), with 0.02\% not specified (n=3). Participants were relatively familiar with AI, with an average of 14.3 out of 25 on overall AI literacy (median=14.41, SD=4.50, range=5--25).

Participants were generally experienced in team projects at school, having participated in an average of 12 team projects (median=5, SD=25.8, range=0--300). Participants held a relatively positive attitude when asked to rate their overall experience on a scale of 1-Extremely negative to 5-Extremely positive, with an average rating of 3.7 (median=4, SD=0.88). Eight participants didn't report given they had not been involved in any team projects.

Given that personality could affect students' perception of AI after personality misrepresentations, we also provide an overview of the participants' personality (on a scale from 1 to 5): the average of participants' Extroversion is 2.93 (median=2.88, SD=0.86, range=1--5), the average of participants' Agreeableness is 3.86 (median=3.88, SD=0.66, range=2.22--5), the average of participants' Conscientiousness is 3.79 (median=3.78, SD=0.73, range=1.67--5), the average of participants' Neuroticism is 2.83 (median=2.88, SD=0.99, range=1--5), and the average of participants' Openness is 3.76 (median=3.8, SD=0.65, range=1.6--5)

\section{Study 2: Findings Tables and Figures}
\subsection{Examining the Effect of AI Misrepresentation on Changes in Students' Perceptions of AI} \label{appendix:s2effectonchanges}
Students' baseline perceptions of SAMI showed that participants in both conditions had similar initial perceptions in terms of overall trust (accurate condition median=2.33, SD = 1.04; inaccurate condition median=2, SD = 1.02), intelligence (accurate condition median=3.4, SD=1.09; inaccurate condition median=3, SD=0.95), anthropomorphism (accurate condition median=2.2, SD=1; inaccurate condition median=2.2, SD=0.83), and likeability(accurate condition median=3.2, SD=1.01; inaccurate condition median=3, SD=0.79). 

\begin{figure}[h]
    \centering
        \begin{subfigure}[b]{0.24\textwidth}
            \centering
            \includegraphics[width=\textwidth]{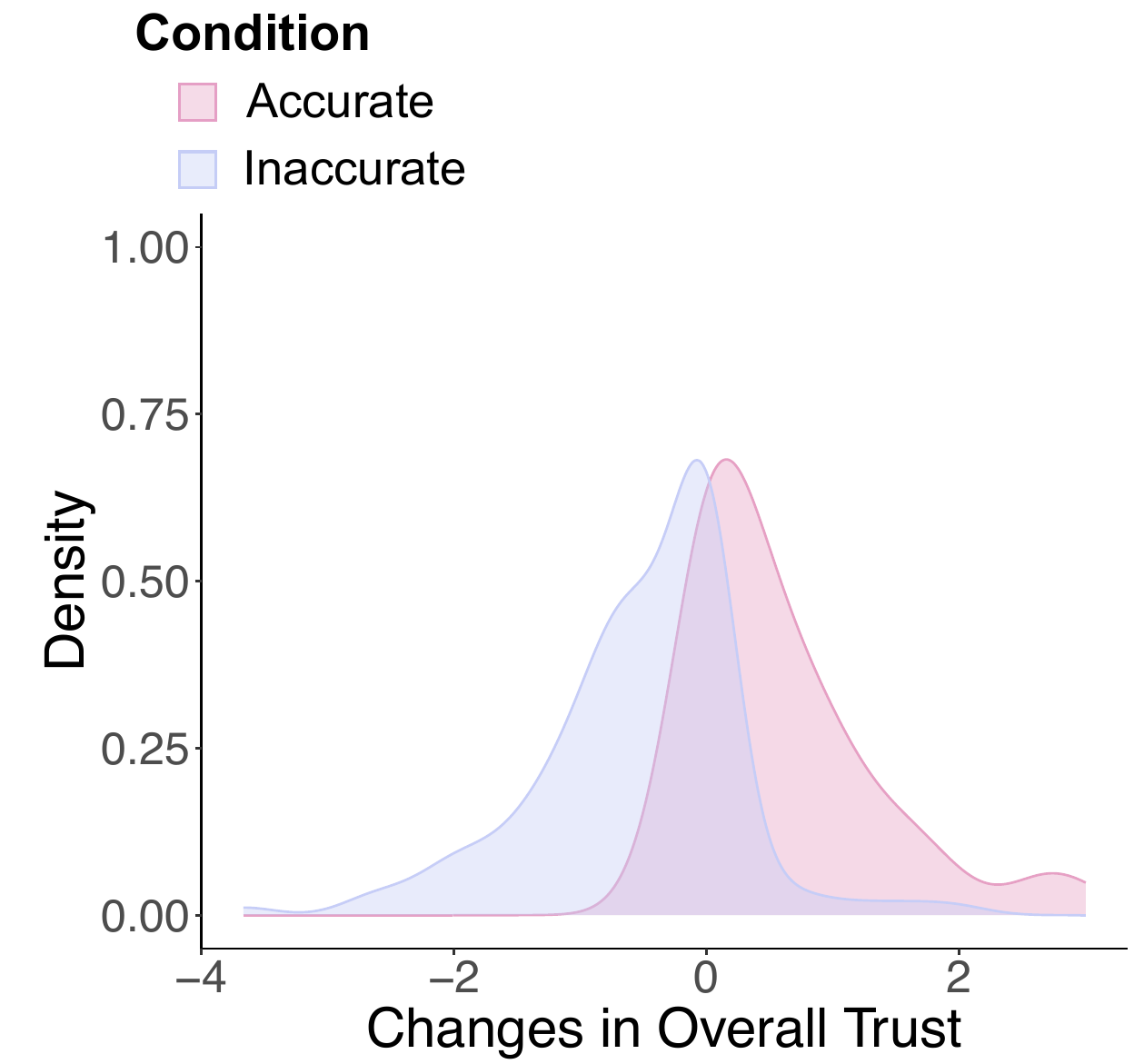}
            \caption{}
            \label{fig:density_trust}
        \end{subfigure}
        \hfill
        \begin{subfigure}[b]{0.24\textwidth}
            \centering
            \includegraphics[width=\textwidth]{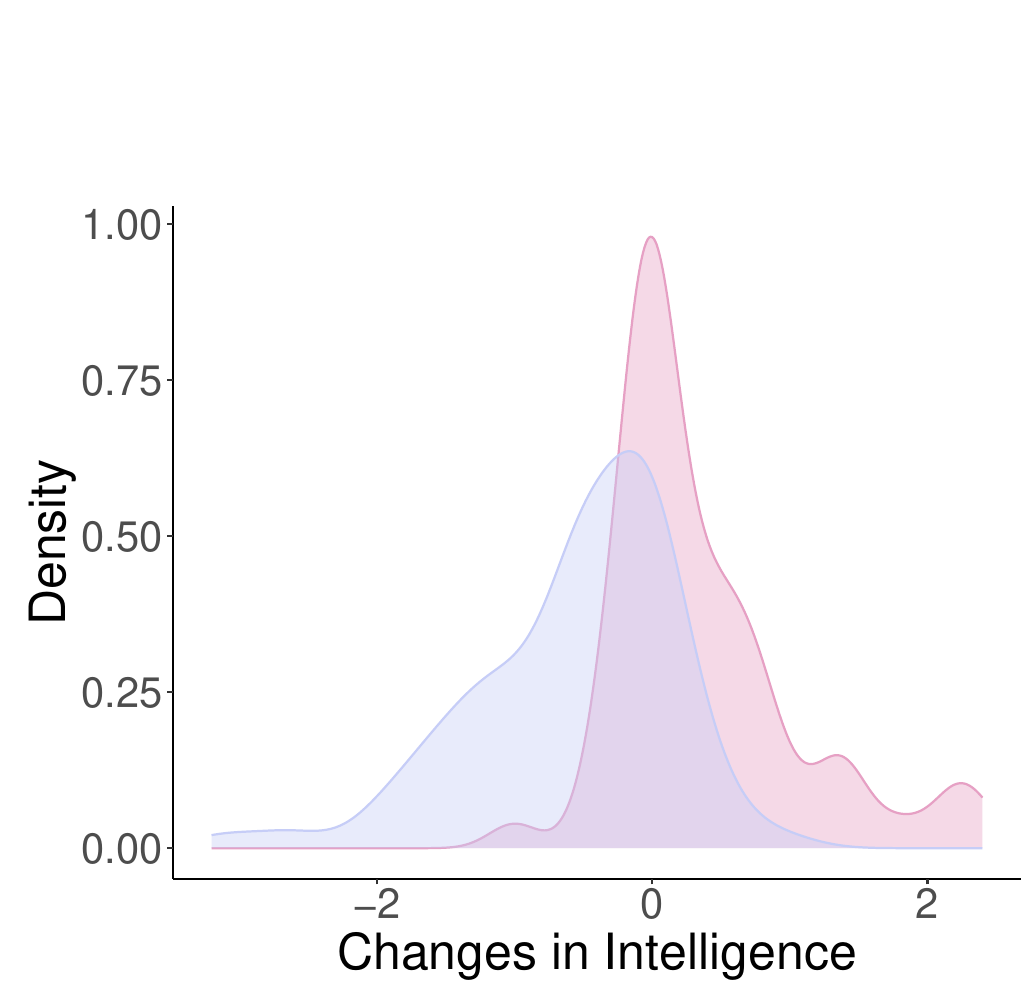}
            \caption{}
            \label{fig:density_int}
        \end{subfigure}
        \hfill
        \begin{subfigure}[b]{0.24\textwidth}
            \centering
            \includegraphics[width=\textwidth]{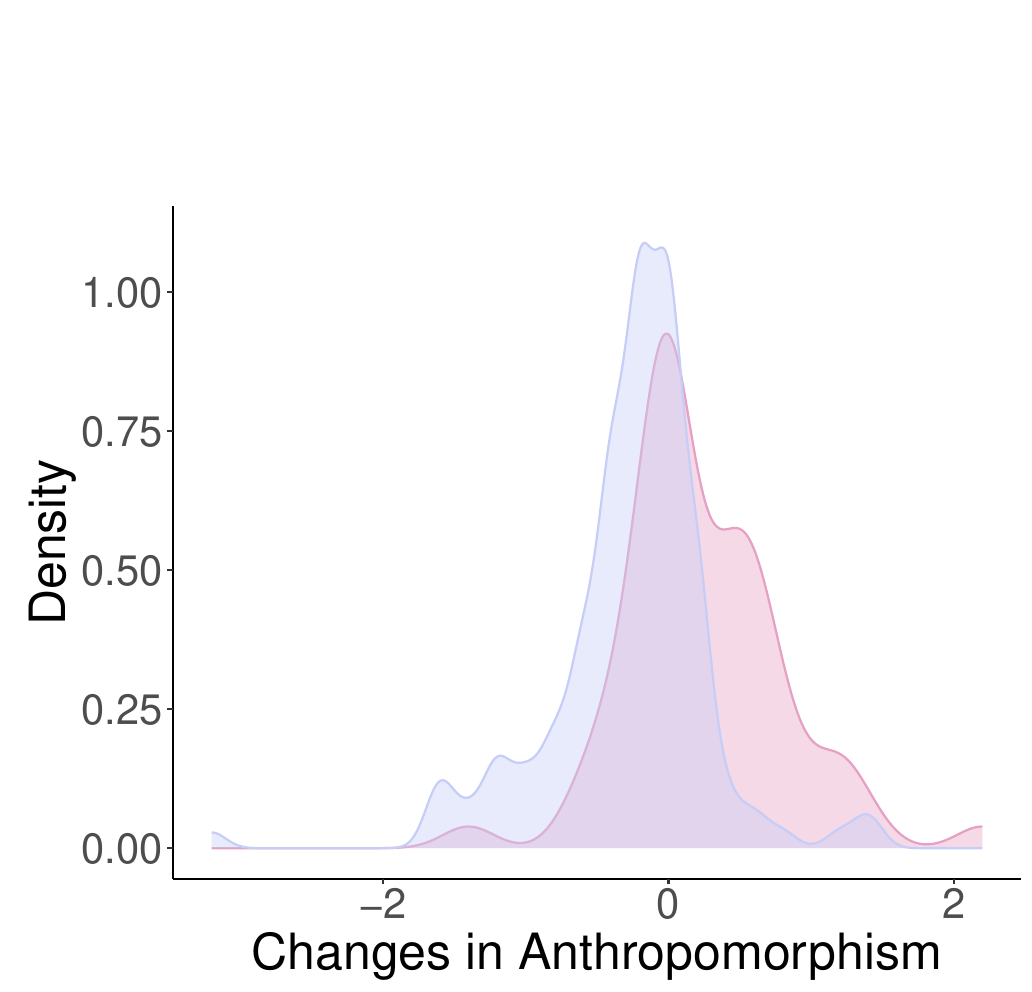}
            \caption{}
            \label{fig:density_anth}
        \end{subfigure}
        \hfill
        \begin{subfigure}[b]{0.24\textwidth}
            \centering
            \includegraphics[width=\textwidth]{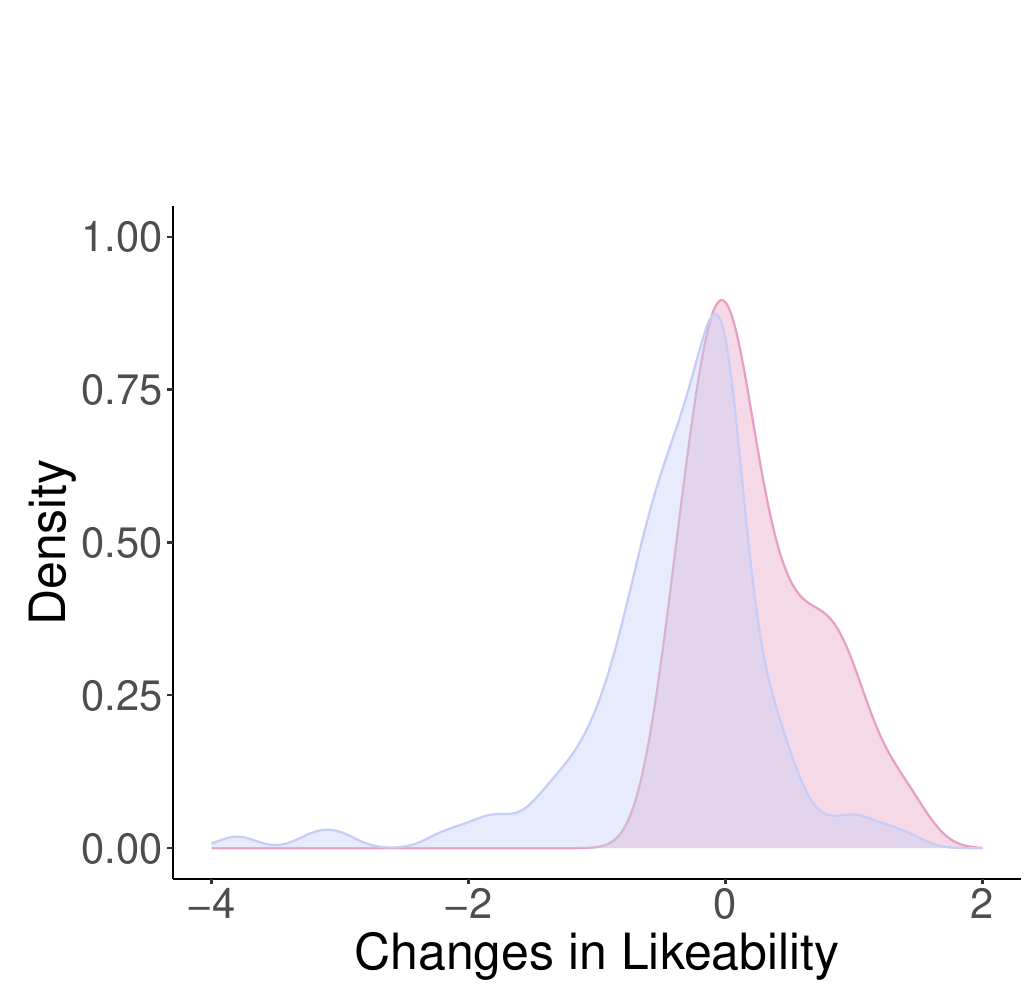}
            \caption{}
            \label{fig:density_like}
        \end{subfigure}
    \caption{Density plots visualizing the participant distribution of changes in overall trust, intelligence, anthropomorphism, and likeability in the accurate and inaccurate conditions.}
    \label{fig:perception_changes_density}
\end{figure}

We then looked at the changes in students' experiment perceptions after we introduced SAMI's inferences in the two conditions. We plotted out the changes in each perception outcome in density plots as shown in Fig.~\ref{fig:perception_changes_density}. We noticed that many participants in the inaccurate condition decreased their perceptions of SAMI after encountering AI misrepresentations compared to participants in the accurate condition. Therefore, encountering AI misrepresentations seemed to have a negative effect on changes in students' perceptions of SAMI. 

We then built four linear regression models with changes in each perception as the outcome to further examine the effect of encountering AI misrepresentation on students' perceptions of SAMI. Our models (based on Equation~\ref{eq1}) show that after controlling for demographics, team project experiences and attitude, as well as the personality dimensions, encountering AI misrepresentation has a significant effect on changes in students' perceptions of SAMI in terms of overall trust, perceived intelligence, anthropomorphism, and likeability (Table~\ref{tab:lm_perceptionchanges}).

\begin{table}[h]
    \centering
    \sffamily
    \footnotesize
    \caption{Results of our regression models(Equation~\ref{eq1}) show that participants in the inaccurate condition had a significant decline in overall trust, perceived intelligence, anthropomorphism, and likeability. The only significant covariate, the Openness personality dimension, is reported in the table. *** p<0.001 ** p<0.01 * p<0.05 . p<0.1}
    \setlength{\tabcolsep}{1.1em}
    \renewcommand{\arraystretch}{1.2}
    \begin{tabular}{lcccccccccccc}
        &   \multicolumn{3}{c}{\textbf{Overall Trust}}   &   \multicolumn{3}{c}{\textbf{Intelligence}}    &   \multicolumn{3}{c}{\textbf{Anthropomorphism}}    &   \multicolumn{3}{c}{\textbf{Likeability}} \\ 
        &   Est.    &   &   S.E &   Est.    &   &   S.E &   Est.    &   &   S.E     &   Est.    &   &   S.E \\ \toprule
        (Intercept) &   0.84    &   &   0.782   &   0.64   &   &   0.728   &   -0.06   &   &   0.578   &   1.39    & * &   0.664 \\
        Openness    &   -0.08   &   &   0.100   &   -0.17   & . &  0.093   &    -0.09   &   &   0.074   &   -0.18   & * &   0.085 \\ \midrule
        Condition (Inaccurate)   &   -1.14   &***&   0.133   &   -0.92    &***&  0.124   &    -0.50   &***&   0.098   &   -0.58   &***&   0.113 \\ 
        &   \multicolumn{3}{c}{Adj. $R^2$=0.283***} &   \multicolumn{3}{c}{Adj. $R^2$=0.234***}     &   \multicolumn{3}{c}{Adj. $R^2$=0.111***}     &   \multicolumn{3}{c}{Adj. $R^2$=0.139***} \\ 
        \bottomrule
    \end{tabular}
    \label{tab:lm_perceptionchanges}
\end{table}

\subsection{AI Literacy as a Moderator} \label{appendix:s2_literacy}
We performed two versions of each of the four models to examine AI literacy's moderating effect (Table~\ref{tab:moderation}).

\begin{table}[h!]
    \centering
    \sffamily
    \footnotesize
    \caption{Results of our regression models with AI literacy in base models (Equation~\ref{eq2}) and base + interaction models (Equation~\ref{eq3}). Results suggested a significant interaction effect between condition and AI literacy in changes in overall trust after encountering AI misrepresentations. However, a significant interaction effect is not found in changes in intelligence, anthropomorphism, and likeability *** p<0.001 ** p<0.01 * p<0.05 . p<0.1}
    \setlength{\tabcolsep}{1em}
    \renewcommand{\arraystretch}{1.2}
    \begin{tabular}{lllllllllllll}
        &   \multicolumn{6}{c}{\textbf{Overall Trust}}   &   \multicolumn{6}{c}{\textbf{Intelligence}}  \\ 
        &   \multicolumn{3}{c}{\textbf{1a. Base}}   &    \multicolumn{3}{c}{\textbf{1b. Base + Interaction}}    &   \multicolumn{3}{c}{\textbf{2a. Base}}   &    \multicolumn{3}{c}{\textbf{2b. Base + Interaction}}  \\ 
        &   Est.    & &     S.E     &   Est.    & &     S.E     &   Est.    & &     S.E     &   Est.    & &     S.E  \\ \toprule
        (Intercept) &   1.01    &**&    0.351   &   0.53    & & 0.416   &   0.79    &*& 0.321   &   0.50    & &  0.384   \\
        Openness    &   -0.06   &  &    0.088   &   -0.07  & &  0.088   &   -0.15   &.& 0.081   &   -0.16   &*& 0.081 \\ \midrule
        Condition (Inacc)    &   -1.12   &***&    0.124   &   -0.33  & &  0.393   &   -0.91   &***& 0.113   &   -0.33   & & 0.363 \\
        AI Literacy    &   -0.01   &  &    0.013   &   0.024  & &  0.022   &   0.01   &  & 0.012   &   0.03   & & 0.020 \\ 
        Condition (Inacc) X AI Literacy    &      &  &       &   -0.06  &*&  0.026   &      &   &    &   -0.03   & & 0.024 \\ 
         &   \multicolumn{3}{c}{Adj. $R^2$=0.296***} &   \multicolumn{3}{c}{Adj. $R^2$=0.309***}     &   \multicolumn{3}{c}{Adj. $R^2$=0.260***}     &   \multicolumn{3}{c}{Adj. $R^2$=0.264***} \\ 
        \bottomrule
    \end{tabular}

    \vspace{3mm}
    
    \setlength{\tabcolsep}{1em}
    \renewcommand{\arraystretch}{1.2}
    \begin{tabular}{lllllllllllll}
        &   \multicolumn{6}{c}{\textbf{Anthropomorphism}}   &   \multicolumn{6}{c}{\textbf{Likeability}}  \\ 
        &   \multicolumn{3}{c}{\textbf{3a. Base}}   &    \multicolumn{3}{c}{\textbf{3b. Base + Interaction}}    &   \multicolumn{3}{c}{\textbf{4a. Base}}   &    \multicolumn{3}{c}{\textbf{4b. Base + Interaction}}  \\ 
        &   Est.    & &     S.E     &   Est.    & &     S.E     &   Est.    & &     S.E     &   Est.    & &     S.E  \\ \toprule
        (Intercept) &   0.54    &*&    0.256   &   0.64    &*& 0.307   &   0.81    &**& 0.296   &   0.63    &.&  0.354   \\
        Openness &   -0.08    & &    0.064   &   -0.08    & & 0.065   &   -0.17    &*& 0.074   &   -0.17    &*&  0.075   \\ \midrule
        Condition (Inacc) &   -0.50    &***&    0.090   &   -0.66    &*& 0.290   &   -0.59    &***& 0.104   &   -0.29    & &  0.334   \\
        AI Literacy &   0.00    & &    0.009   &   -0.01    & & 0.016   &   0.00    & & 0.011   &   0.017    & &  0.018   \\
        Condition (Inacc) X AI Literacy &       & &       &   0.01    & & 0.019   &       & &    &   -0.02    & &  0.022   \\
        &   \multicolumn{3}{c}{Adj. $R^2$=0.137***} &   \multicolumn{3}{c}{Adj. $R^2$=0.134***}     &   \multicolumn{3}{c}{Adj. $R^2$=0.160***}     &   \multicolumn{3}{c}{Adj. $R^2$=0.159***} \\ 
        \bottomrule
    \end{tabular}
    \label{tab:moderation}
\end{table}

\subsection{Post-hoc analysis} \label{appendix:s2posthoc}
\begin{figure}[h!]
    \centering
        \begin{subfigure}[b]{0.3\textwidth}
            \centering
            \includegraphics[width=\textwidth]{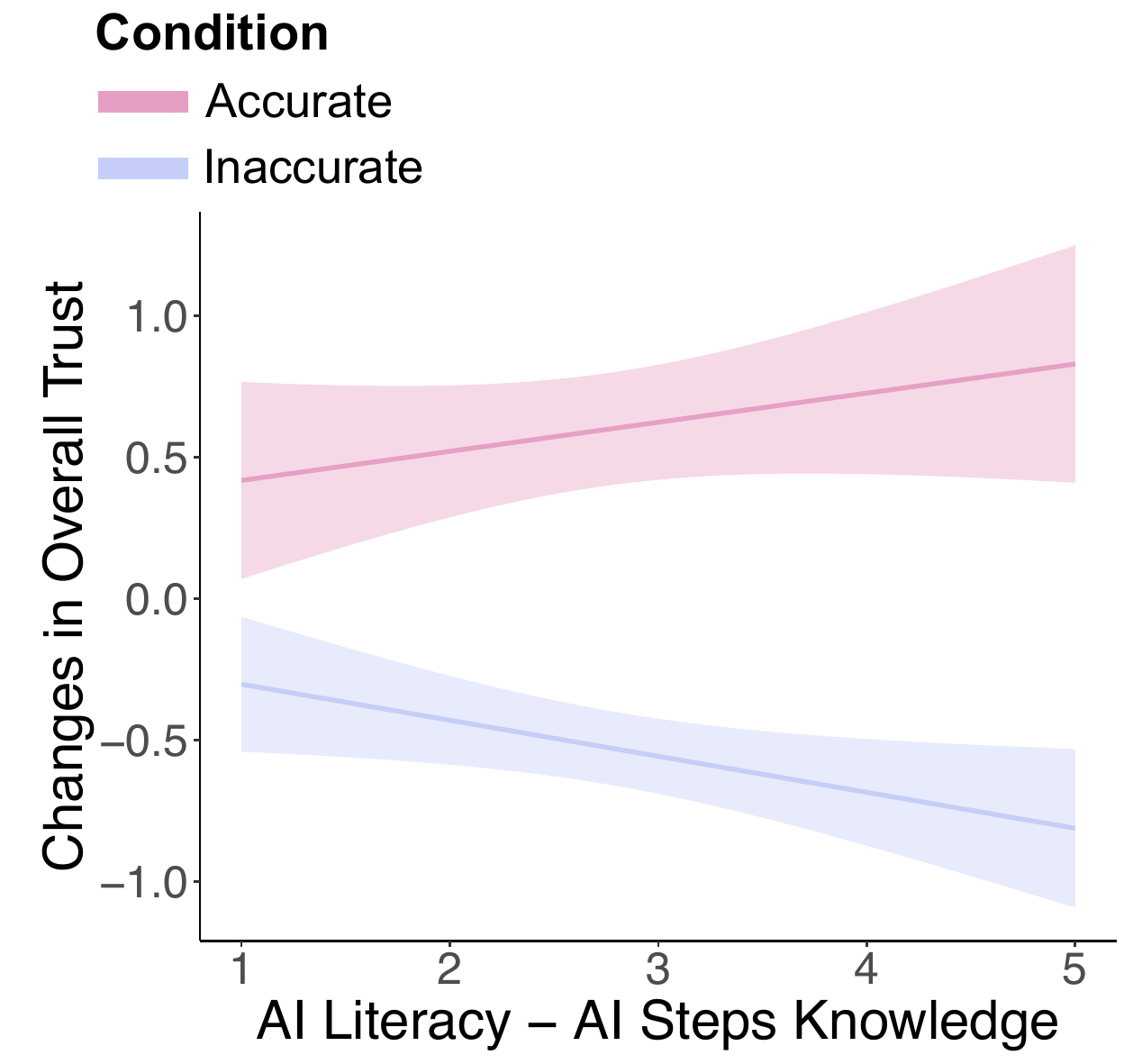}
            \caption{}
            \label{fig:trust_steps}
        \end{subfigure}
        \hfill
        \begin{subfigure}[b]{0.3\textwidth}
            \centering
            \includegraphics[width=\textwidth]{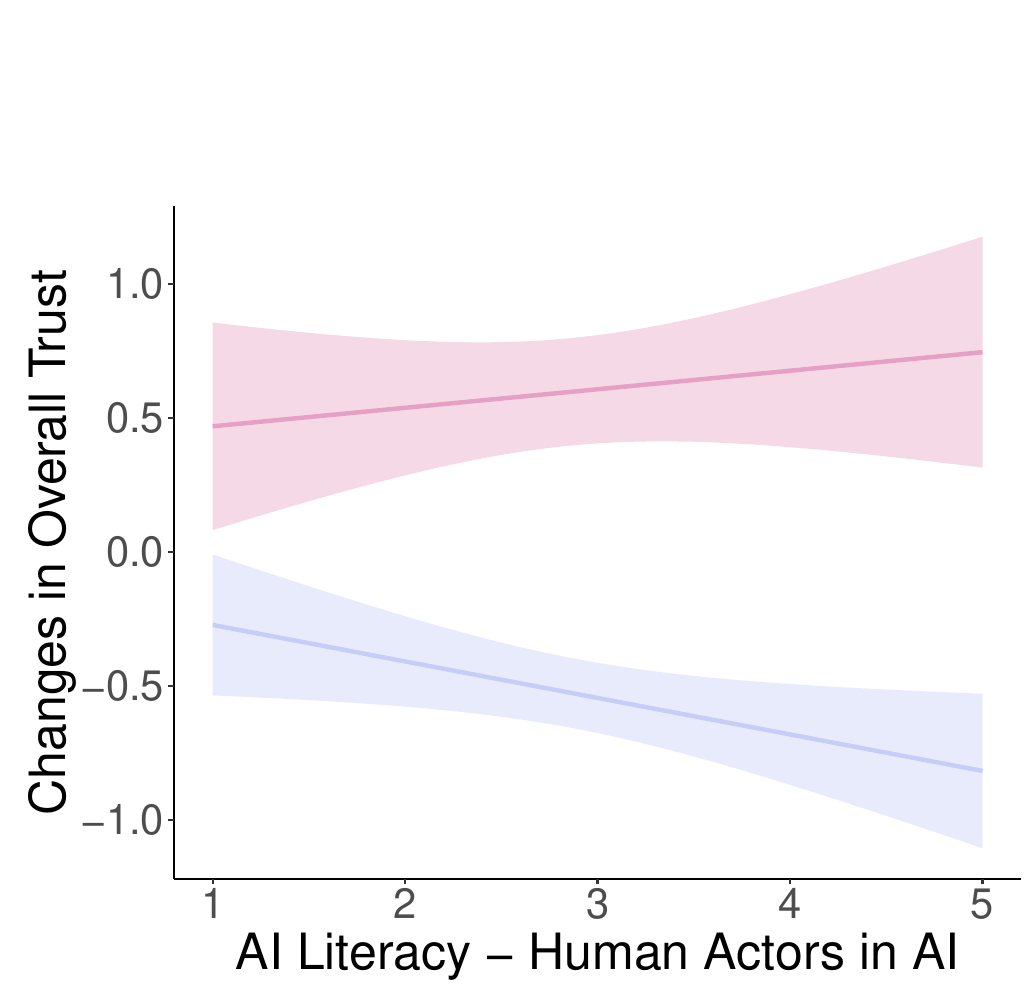}
            \caption{}
            \label{fig:trust_human}
        \end{subfigure}
        \hfill
        \begin{subfigure}[b]{0.3\textwidth}
            \centering
            \includegraphics[width=\textwidth]{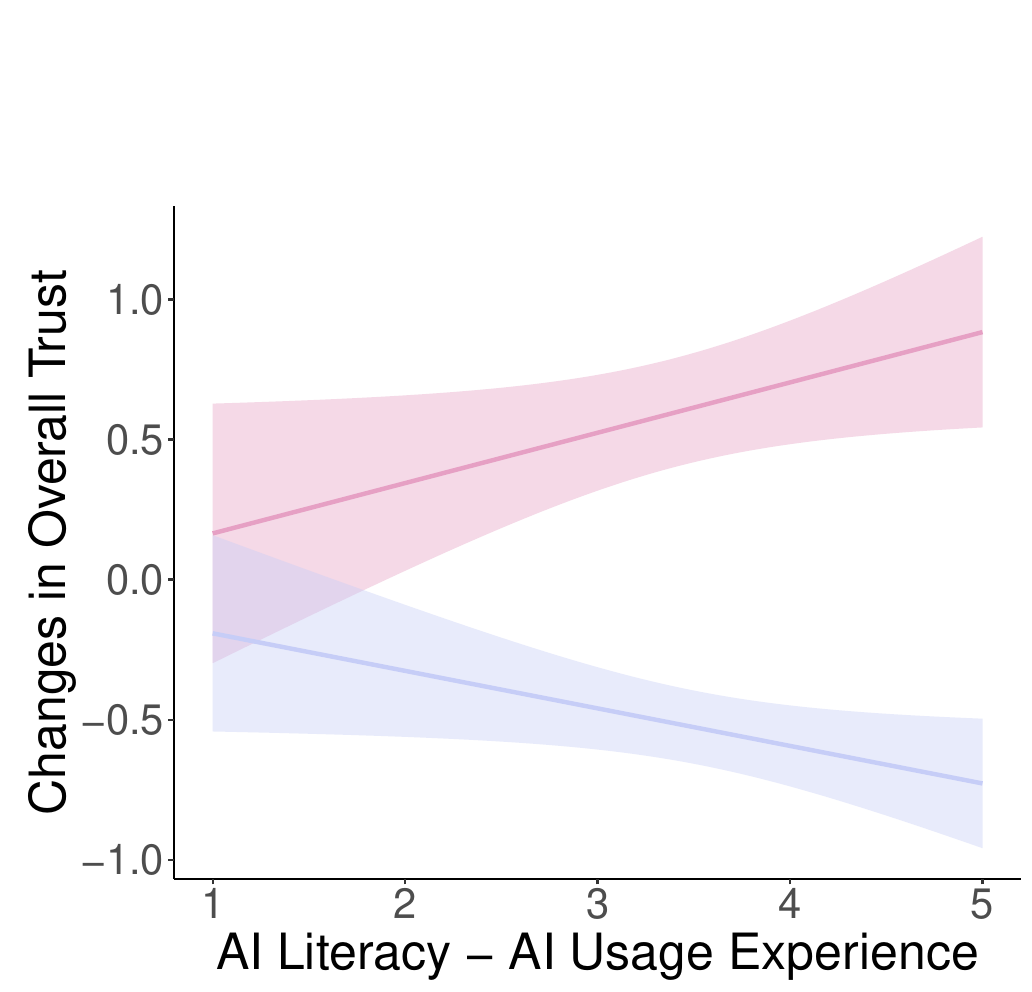}
            \caption{}
            \label{fig:trust_use}
        \end{subfigure}
    \caption{Post-hoc analysis with three general AI literacy sub-dimensions: AI steps knowledge, human actors in AI knowledge, AI usage experience. A significant effect between each literacy sub-dimension and condition was found in all three linear regression models with changes in overall trust as the outcome variable. All three models controlled for the Openness personality dimension. We also found a significant main effect of AI usage experience on students' changes in overall trust of SAMI after encountering AI misrepresentation.}
    \label{fig:post-hoc_analysis}
\end{figure}

In our post-hoc analysis, we ran five linear regression models with students' changes in overall trust in SAMI as the outcome variable and included an effect of condition, each AI literacy dimension, and an interaction effect between condition and each AI literacy dimension. We controlled for the Openness personality dimension in all five models. Among the five individual models, we found a significant interaction effect between condition and students' AI steps knowledge, human actors in AI knowledge, and AI usage experience. We plotted out the interaction effects in Figure~\ref{fig:post-hoc_analysis}.

\clearpage
\section{SAMI Inference Fabrication Rule Book} \label{supp:fabrication_rule}
\noindent \textbf{Steps to fabricate SAMI Inferences:}

\begin{enumerate}
    \item Calculate the personality dimension score for the participant
    \item Choose up to three personality dimensions that are scored either below 2.5 or above 3.5; if these conditions are not satisfied, choose the dimensions that are on the more extreme side (e.g., 2.3 is better than 2.9)--- not 3.0. 
    \item Under each selected personality dimension, select the extreme statements (scored as either 1, 2, 4, or 5) and compose accordingly based on which condition the participant is in (inaccurate or accurate condition). Select a total of 8-10 inferences to compose SAMI’s response.
    \item When fabricating the inferences, please check the table below for how to reverse or paraphrase each original statement.
    \begin{enumerate}
        \item For inaccurate condition, paraphrase the statements that the participant disagreed on (rated as 1 or 2) and/or reverse the statements that the participant agreed on (rated as 4 or 5).
        \item For accurate condition, paraphrase the statements that the participant agreed on (rated as 1 or 2) and/or reverse the statements that the participant disagreed on (rated as 4 or 5).
    \end{enumerate}
    \item Additional rules that we are following:
    \begin{enumerate}
        \item The total inferences SAMI makes should be about 10 inferences in total.
        \item Try to get a mix of positive and negative inferences by following roughly 40\% negative inferences and 60\% positive inferences. In the table, inferences marked with ``[N]'' represents negative inferences, inferences marked with ``[P]'' represents positive inferences
    \end{enumerate}
\end{enumerate}

\textbf{Inferences table can be found in Table~\ref{tab:inference_table}.}

\begin{table}
    \centering
    \small
    \footnotesize
    \caption{Table that listed out the original statements in the Big Five personality inventory, the paraphrase for accurate inferences, and the reverse for inaccurate inferences. }
    \begin{tabular}{p{1.5cm}p{3cm}p{5cm}p{5cm}} \toprule
        \textbf{Personality Dimensions} & \textbf{Original Statements in the Big Five Inventory} & \textbf{Paraphrase}  &   \textbf{Reverse} \\ \midrule
                &   Is talkative    &  You are talkative. [P]     &    You tend to be quiet. [N] \\ 
                & Is reserved  &  You are reserved.[N]    &   You are outgoing and sociable. [P]  \\
                & Is full of energy  &  You are full of energy. [P]    &   You tend to lack energy. [N]  \\ 
               & Generates a lot of enthusiasm  & You generate a lot of enthusiasm. [P]      & You are not readily enthusiastic. [N]  \\
        Extraversion  & Tends to be quiet    & You tend to be quiet. [N]    & You are talkative. [P]   \\
               & Has an assertive personality    & You have an assertive personality. [N]    & You have a low-key personality[P] \\
               & Is sometimes shy, inhibited    & You are sometimes shy and inhibited. [N]    & You are confident and extroverted. [P]   \\
               & Is outgoing, sociable    & You are outgoing and sociable. [P]    & You are reserved. [N]  \\ \midrule
    & Tends to find fault with others    &   You tend to find fault with others. [N]  & You tend to see the good in others. [P]  \\
    & Is helpful and unselfish with others   &  You are helpful and unselfish with others. [P]    &   You are self-centered and unhelpful. [N]  \\
    & Starts quarrels with others   &  You often start quarrels with others. [N]    &  You are good at de-escalating conflicts. [P]  \\
    & Has a forgiving nature   &  You have a forgiving nature. [P]    &   You hold onto people’s mistakes. [N] \\
    Agreeableness & Is generally trusting   &  You are generally trusting. [P]    &   You are cautious about trusting others. [N]  \\
    & Can be cold and aloof   &  You can be cold and aloof. [N]    &   You are warm and friendly. [P]  \\
    & Is considerate and kind to almost everyone   &  You are considerate and kind to almost everyone. [P]    &   You are sometimes rude to others. [N]  \\
    & Is sometimes rude to others   &  You are sometimes rude to others. [N]    &   You are considerate and kind to almost everyone. [P] \\ \midrule
    
        &   Does a thorough job   &   You do a thorough job. [P]  &   You tend to do things in a hurried manner. [N]  \\
        &   Can be somewhat careless    &   You can be somewhat careless. [N]   &   You are careful and meticulous. [P] \\
        &   Is a reliable worker    &   You are a reliable worker. [P]  &   You are not consistently dependable in your work. [N] \\
        &   Tends to be disorganized    &   You tend to be disorganized. [N]    &   You are organized. [P] \\
    Conscientiousness   &   Tends to be lazy &  You tend to be lazy. [N]    &   You are hard-working. [P] \\
        &   Perseveres until the task is finished & You persevere until the task is finished. [P] & You tend to give up on tasks easily. [N] \\
        &   Does things efficiently &   You do things efficiently. [P]  &   You tend to be inefficient in your approach to tasks. [N] \\
        &   Makes plans and follows through with them   &   You make plans and follow through with them. [P]    &   You don’t like rigid plans and structures and avoid planning ahead. [P] \\
        &   Is easily distracted    &   You are easily distracted. [N]  &   You are highly focused. [P] \\ \midrule
        &Is depressed, blue &   You often feel sad and depressed. [N]   &   You are always cheerful and happy. [P] \\
        &   Is relaxed, handles stress well &   You are relaxed and you handle stress well. [P] &   You are tense. [N]  \\
        &   Can be tense    &   You can be tense. [N]   &   You are relaxed and you handle stress well. [P] \\
    Neuroticism &   Worries a lot & You worry a lot. [N]    &   You are carefree. [P] \\
        &   Is emotionally stable, not easily upset     &   You are emotionally stable and not easily upset. [P]    &   You can be moody. [N] \\
        &   Can be moody    &   You can be moody. [N]   &   You are emotionally stable. [P] \\
        &   Remains calm in tense situations    &   You remain calm in tense situations. [P]    &   You get nervous easily. [N] \\
        &   Gets nervous easily &   You get nervous easily. [N] &   You remain calm in tense situations. [P]  \\ \midrule
        &   Is original, comes up with new ideas    &   You are original and you often come up with new ideas. [P]  &   You tend to be unimaginative. [N]   \\
        &   Is curious about many different things  &   You are curious about many different things. [P]    &   You exhibit limited curiosity towards diverse subjects. [N] \\
        &   Is ingenious, a deep thinker    &   You are ingenious and a deep thinker. [P]   &   You don't often exhibit creativity or engage in complex thinking. [N] \\
        &   Has an active imagination   &   You have an active imagination. [P] &   You are not particularly imaginative in your thinking. [N] \\
    Openness &  Is inventive   &    You are inventive. [P]  &   You tend to follow traditional ways. [N] \\
        &   Values artistic, aesthetic experiences &    You value artistic and aesthetic experiences. [P]   &   You typically are not drawn to arts and creative expressions.  [P] \\
        &   Prefers work that is routine    &   You prefer work that is routine. [P]    &   You get bored with the routine and mundane. [P] \\
        &   Likes to reflect, play with ideas   &   You like to reflect and play with ideas. [P]    &   You don’t like spending time playing with ideas. [N] \\
        &   Has few artistic interests  &   You have few artistic interests. [N]    &   You value artistic and aesthetic experiences. [P] \\
        &   Is sophisticated in art, music, or literature & You are sophisticated in art, music, or literature. [P] &   You have limited appreciation of art, music, or literature.  [N] \\
    \bottomrule
    \end{tabular}
    \label{tab:inference_table}
\end{table}

\section{Study 1 Session and Interview Protocol} \label{supp:s1_protocol}
\textbf{Study Introduction} 

[Note that we used a study slide deck to show participants the sample and personal inferences. The study deck also contains slides for study introduction, debriefing form, and other things to go through with the participants during the session. The following paragraphs summarize the main idea of the slide deck for study introduction] 

The study slide deck first shows the motivation of our project, that it is difficult to find teammates for school projects in both in-person classes and online classes. The question that our research team is trying to resolve is, how can we help students to form teams more easily and efficiently? 

Our team built an AI agent named SAMI, which stands for Social Agent Mediated interaction. SAMI is an AI agent that can recommend potential teammates based on its understanding of the student, inferred from students’ self-introduction. Due to the constraint of this study setting, we will only focus on SAMI’s understanding of the student, instead of providing actual team-matching results for the participants.  

The goal of the study is threefold: (1) To assess the perceived accuracy of SAMI’s inferences about the student. (2) To understand how students perceive SAMI’s inferences. (3) To understand how students think of SAMI. 

[Facilitator briefly introduces the session procedures regarding the baseline samples, SAMI’s inference about them, the perception measurements, and the interview portion of the study] 

Reminders: There is no right answer to any of the questions. We are not evaluating you, we are evaluating the AI agent. Be honest and straightforward--- you will not hurt our feelings. 

Do you have any questions for me? 

[If participant did not sign the consent form prior to coming to the study session, ask them to review and sign the consent form before proceeding] \\

\noindent \textbf{Review Inferences and Fill Out Perception Measures}

\begin{figure}[h]
    \centering
    \includegraphics[width=\textwidth]{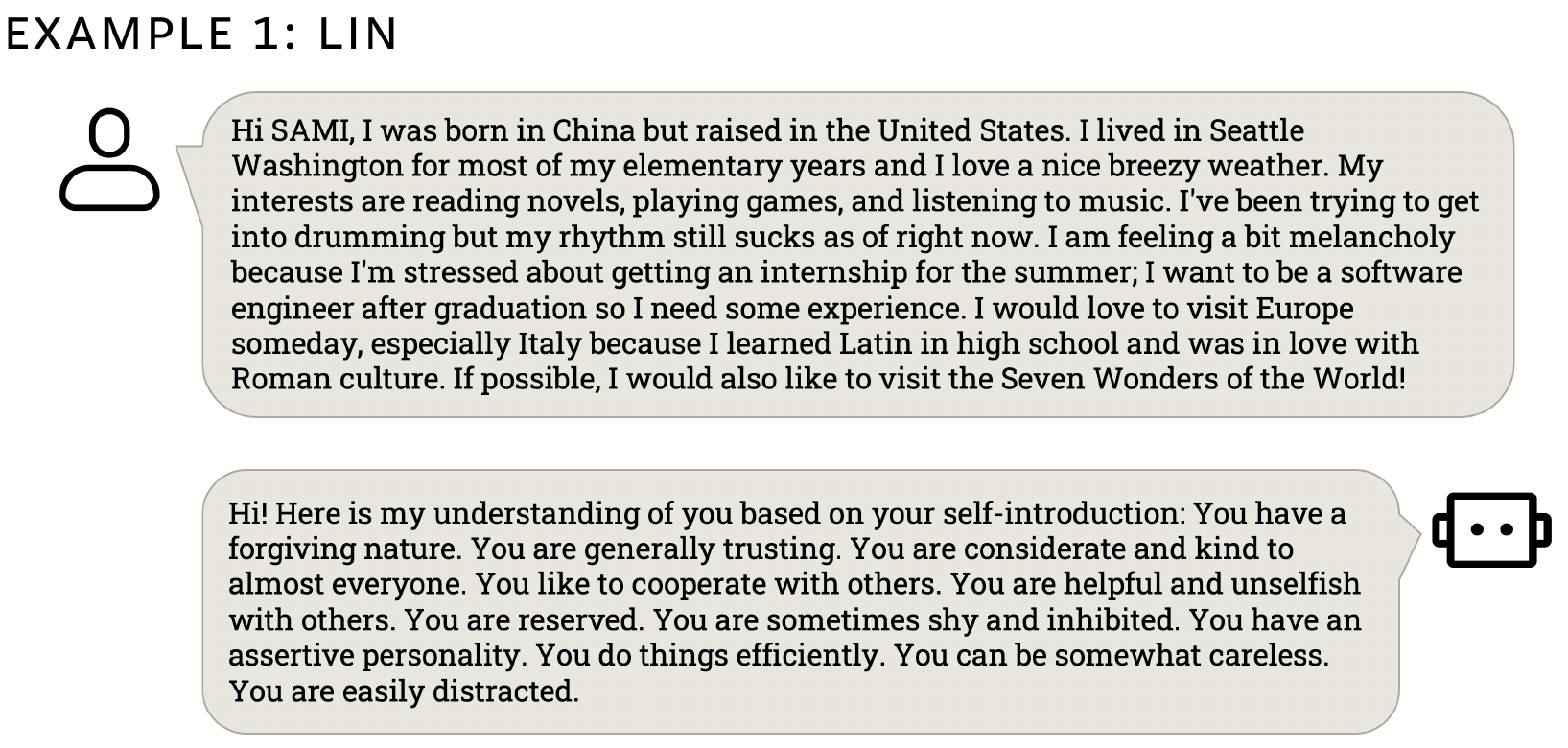}
    \caption{This figure shows the accurate sample about a student named Lin.}
    \label{supp:lin_sample}
\end{figure}

\begin{figure}[h]
    \centering
    \includegraphics[width=\textwidth]{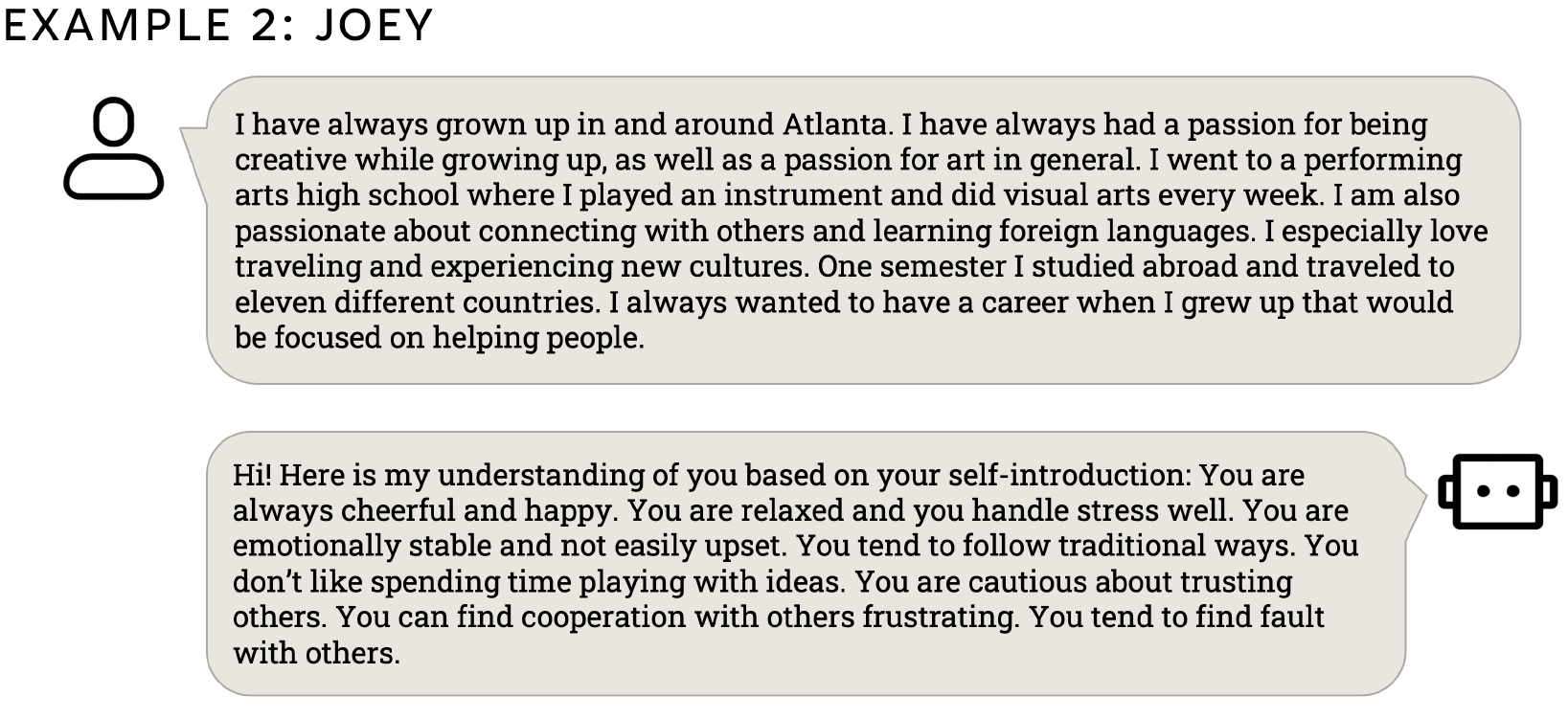}
    \caption{This figure shows the inaccurate sample about a student named Joey.}
    \label{supp:joey_sample}
\end{figure}

Participants were shown the sample of a student named Lin (accurate sample, see Fig.~\ref{supp:lin_sample}) and the sample of a student named Joey (inaccurate sample, see Fig.~\ref{supp:joey_sample}). The samples were shown one after the other and the display order was randomized. After participants reviewed these two samples, they were directed to the Qualtrics survey to record their baseline perceptions of SAMI.

Participants were then shown their own self-introduction and SAMI's inferences about them. After that, participants went back to the Qualtrics survey to record their experiment perceptions of SAMI. Note that the measurements for baseline perceptions and experiment perceptions are the same set of measurements (see Supplementary Materials~\ref{supp:s1_exp_survey}). After participants submitted the Qualtrics survey after recording their experiment perceptions, the facilitator proceeded with the semi-structured interview. \\

\noindent \textbf{Semi-Structured Interview}

Note that this is a protocol for semi-structured interview, hence not all questions asked during the interview is covered in this protocol, and not all questions listed here were asked during the interview. This protocol only acts as a guide for the interviews and the session moderator has the flexibility to ask questions within this protocol or outside of this protocol. 

[Start recording]
Let’s walk through your experience for each pair of introduction and SAMI inferences. [Pull up participants’ survey responses on the side.]

\begin{enumerate}
    \item Sample 1
    \begin{enumerate}
        \item What was your first impression of SAMI?
        \item Anything unexpected or surprising about SAMI’s response?
        \item How do you feel about SAMI's inferences for sample 1?
        \item How do you feel about SAMI after seeing sample 1? [Go through participants' survey responses after seeing sample 1] Anything that didn't get captured by the survey questions?
        \item At this point, how did you think SAMI work? How did you think SAMI made each inference?
    \end{enumerate}
    \item Sample 2
    \begin{enumerate}
        \item How do you feel about SAMI's inferences for sample 2?
        \item What was your impression of SAMI after you saw sample 2? Anything changed?
        \item How do you feel about SAMI after seeing sample 2? [Go through participant's survey responses after seeing sample 2] Anything that didn't get captured by the survey questions?
        \item After seeing sample 2, do you feel like you have a better or worse understanding of how SAMI works?
    \end{enumerate}
    \item SAMI's inference about the participant
    \begin{enumerate}
        \item Now that you've seen SAMI's inferences about you, what do you think of SAMI in general?
        \begin{enumerate}
            \item How do you feel about SAMI's response?
            \item Anything unexpected or surprising about SAMI's response?
            \item What do you like or dislike about SAMI's response?
        \end{enumerate}
        \item Could you walk through SAMI's inferences line by line and let me know what you think of each one?
        \begin{enumerate}
            \item How accurate do you think these inferences are? Why or why not? How did you draw that conclusion?
            \item How do you feel about each inference?
            \item How do you think SAMI made each inference?
        \end{enumerate}
        \item Did you learn anything surprising about yourself from SAMI's inferences?
        \item How do you think your perception about SAMI changed, if any, after seeing SAMI's inferences about you? [Go through participant's survey responses after seeing their own inferences]
        \item What about your understanding of how SAMI works? Do you have a better or worse understanding of how SAMI works now that you saw your own inferences?
        \item After seeing SAMI's inferences about you, if you were to modify your introduction and let SAMI make inferences again, how would you change your introduction, and why?
        \item How do you think SAMI can be improved? Are there anything that you wish you had known about SAMI before seeing SAMI's inferences about you to better prepare you for this response?
        \item If we were to develop SAMI as a chatbot or a conversational agent, how do you envision the conversation would continue from here ideally? What would you say to SAMI next?
        \item How do you feel about team matching for school projects based on these inferences?
        \begin{enumerate}
            \item What additional inferences do you think would help for team matching?
            \item Who do you think should have access to these inferences? Teachers, TAs, classmates, teammates?
            \item In this study SAMI drew these inferences from your self-introduction, are there other kinds of data that you are willing to offer to SAMI to make better inferences about you? GPA, class history, skillsets, professional experience, social media data?
        \end{enumerate}
    \end{enumerate}
\end{enumerate}

\noindent \textbf{Debriefing}

Thank you so much for completing this study! Now that you have completed the study, I want to share more about the study with you. 

Here is the debriefing form and I will go through it with you. [Read and show the debriefing form on the slide]

\underline{Debriefing Form:} The real purpose of this study is to understand the impact of AI mistakes on users’ perceptions of AI, and how users could identify, react, and be better prepared to deal with AI mistakes during human-AI interactions, instead of assessing SAMI’s inference accuracy like I told you since the beginning of this study. 
 
For the purpose of making SAMI’s capability more advanced and cutting-edge like existing AI system, we led you to believe that SAMI could make implicit inferences such as your personality during this study. That was not true--- SAMI does not have the capability of making implicit inferences like personality based on a paragraph of self-introduction. All the SAMI inferences were generated by human researchers manually instead of by SAMI.

In this study, participants were randomly assigned to either receive accurate SAMI inferences or inaccurate SAMI inferences for their personalized response so that we can better compare and contrast students’ reactions in these two conditions. All the inferences SAMI generated for Sample 1, 2, and your personalized response were all generated by human researchers based on the personality test results, not based on the self-introductions. SAMI’s inferences about you was based on the personality test that you filled out during the screening survey. Depending on which condition you were randomly assigned in, SAMI’s response was generated intentionally to be accurate or inaccurate. \\

Do you have any questions, comments, or thoughts after knowing this? Do you think you are in the inaccurate condition or accurate condition? This is how we fabricated your inferences…

Thanks again for participating in the study! The \$25 gift card will be sent to your email within 2 days. Please let us know if you don’t receive it by then!

\section{Study 1 Preliminary Survey} \label{supp:s1_prelim_survey}
\begin{enumerate}
    \item Below are a number of characteristics that may or may not apply to you. For example, do you agree that you are someone who likes to spend time with others? Please rate each statement on a scale of 1 to 5 (1-Disagree Strongly; 2-Disagree a little; 3-Neither agree nor disagree; 4-Agree a little; 5-Agree strongly) to indicate the extent to which you agree or disagree with that statement. 

    I see myself as someone who... 
    \begin{itemize}
        \item Is talkative
        \item Tends to find fault with others
        \item Does a thorough job
        \item Is depressed, blue
        \item Is original, comes up with new ideas
        \item Is reserved
        \item Is helpful and unselfish with others
        \item Can be somewhat careless
        \item Is relaxed, handles stress well
        \item Is curious about many different things
        \item Is full of energy
        \item Starts quarrels with others
        \item Is a reliable worker
        \item Can be tense
        \item Is ingenious, a deep thinker
        \item Generates a lot of enthusiasm
        \item Has a forgiving nature
        \item Tends to be disorganized
        \item Worries a lot
        \item Has an active imagination
        \item Tends to be quiet
        \item Is generally trusting
        \item Tends to be lazy
        \item Is emotionally stable, not easily upset
        \item Is inventive
        \item Has an assertive personality
        \item Can be cold and aloof
        \item Perseveres until the task is finished
        \item Can be moody
        \item Values artistic, aesthetic experiences
        \item Is sometimes shy, inhibited
        \item Is considerate and kind to almost everyone
        \item Does things efficiently
        \item Remains calm in tense situations
        \item Prefers work that is routine
        \item Is outgoing, sociable
        \item Is sometimes rude to others
        \item Makes plans and follows through with them
        \item Gets nervous easily
        \item Likes to reflect, play with ideas
        \item Has few artistic interests
        \item Likes to cooperate with others
        \item Is easily distracted
        \item Is sophisticated in art, music, or literature
    \end{itemize}
    \item (Open-ended question) SAMI's teammate recommendation will be based on inferences made about the students from their self-introduction. Imagine that you would like SAMI to recommend potential teammates to you for a school project (e.g., extended class project, final-year team project), please write a paragraph (at least 6 sentences)  to introduce yourself to SAMI. 
    
    Consider this a free-flowing essay, where you write different things about you (e.g., where you grew up, your interests and hobbies, your feelings, thoughts, and emotions, your dreams and passions, you career goals, fun facts about you) and let your thoughts flow freely in your writing without pauses. The more you write, the better it will help us during the study!
    \item On a scale of 1 to 5, how would you rate your current technological expertise? For the purposes of this survey, we’re primarily concerned with your computer and web-based skills. We’ve defined three points on the scale as follows. These tasks represent some of the things a person at each level might do.
    \begin{enumerate}
        \item Beginner (characterized as 1 and 2 on scale): Able to use a mouse and keyboard, create a simple document, send and receive e-mail, and/or access web pages 
        \item Intermediate (characterized as 3 on scale): Able to format documents using styles or templates, use spreadsheets for custom calculations and charts, and/or use graphics/web publishing 
        \item Expert (characterized as 4 and 5 on scale): Able to use macros in programs to speed tasks, configure operating system features, create a program using a programming language, and/or develop a database.
    \end{enumerate}
    \item On a scale of 1 to 5, how would you rate your general attitude towards AI technology (e.g., shopping/music recommendation algorithm, chatbot, etc.)
    \begin{itemize}
        \item 1 – Very Negative: You don’t find AI technology useful at all and have very little trust in AI technology. Using AI technology also elicits negative emotion from your (e.g., anxiety, stress, anger)
        \item 2 – Neutral to Negative
        \item 3 - Neutral: You don’t have any strong positive or strong negative feelings towards AI technology.
        \item 4 - Neutral to Positive
        \item 5 - Very Positive: You trust AI technology a lot and find it significantly improved your everyday life. Using AI technology also elicits positive emotion from you (e.g., joy, satisfaction, relaxation)
    \end{itemize}
    \item If needed, please feel free to elaborate or provide more context on your previous answer regarding your attitudes towards AI technology.
    \item What is your name?
    \item How old are you? (Please enter a number)
    \item What is your gender?
    \begin{itemize}
        \item Woman
        \item Man
        \item Non-binary
        \item Prefer not to disclose
        \item Prefer to self-describe: 
    \end{itemize}
    \item What is your current level of study?
    \begin{itemize}
        \item Undergraduate
        \item Master
        \item Doctorate
    \end{itemize}
    \item What major(s) are you in?
    \item What is your academic or professional background? (If applicable)
    \item Why are you interested in participating in our study?
    \item Which email address should we reach you at for further scheduling?
\end{enumerate}

\section{Study 1 Experiment Survey} \label{supp:s1_exp_survey}
\textit{Note that this survey measurement was used three times during the user study session. Once to measure students' perceptions after seeing sample 1, once after seeing sample 2, and a third time after participants saw SAMI's inferences about them.}
\begin{enumerate}
    \item What is your participant ID? \\
    
    The following questions are meant to capture your perceptions of SAMI after seeing Sample 1:
    \item On a scale of 1 to 5, how would you rate the accuracy of SAMI’s inferences about the student in Sample 1?
    \begin{itemize}
        \item 1- Not accurate at all
        \item 2
        \item 3- Somewhat accurate
        \item 4
        \item 5- Very accurate
    \end{itemize}
    \item Now that you have seen an example of how SAMI works, please answer the following questionnaire based on your current impression of SAMI. Please rate each of the following statement based on how much you agree with it on a scale of 1 to 5, 1 indicates strongly disagree and 5 indicates strongly agree.
    \begin{itemize}
        \item I believe that there could be negative consequences when using SAMI. 
        \item I feel I must be cautious when using SAMI.
        \item It is risky to interact with SAMI.
        \item I believe that SAMI will act in my best interest.
        \item I believe that SAMI will do its best to help me if I need help.
        \item I believe that SAMI is interested in understanding my needs and preferences.
        \item I think that SAMI is competent and effective in making accurate inferences about me as a person.
        \item I think that SAMI performs its role as a social recommendation agent very well.
        \item I believe that SAMI has all the functionalities I would expect from a social recommendation agent. 
        \item If I use SAMI, i think I would be able to depend on it completely.
        \item I can always rely on SAMI for recommending students that fit my social needs.
        \item I can trust the information presented to me by SAMI. 
    \end{itemize}
    \item Please answer the following questionnaire based on your current impression of SAMI. The following questions will give you a spectrum from one quality to the other on a scale of 1 to 5, such as from “Unkind (1)” to “Kind (5).” Please rate your perception of SAMI along each of these spectrum:
    \begin{figure}[h]
        \centering
        \includegraphics[width=0.5\textwidth]{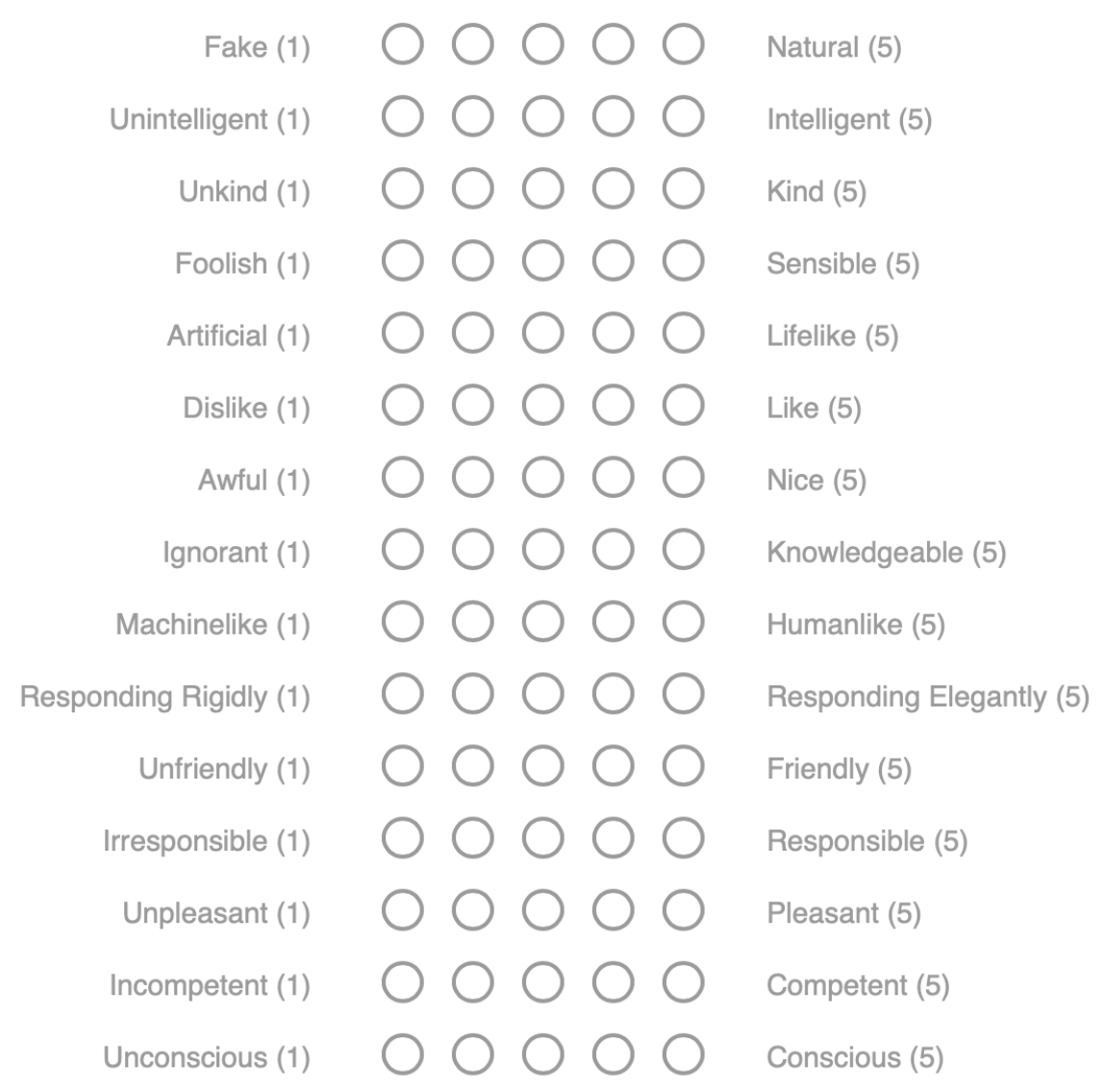}
        \caption{This figure shows the Godspeed questionnaire we used to measure students' social perceptions of SAMI after seeing the samples.}
        \label{supp:godspeed_survey}
    \end{figure}
\end{enumerate}

\section{Study 2 Preliminary Survey} \label{supp:s2_prelim_survey}
(Insert study consent form here)
\begin{enumerate}
    \item Do you consent to participate in this study?
    \begin{itemize}
        \item Yes, I agree to participate.
        \item No, I do not want to participate. 
    \end{itemize}
    \item (Open-ended question) SAMI's teammate recommendation will be based on inferences made about the students from their self-introduction. Imagine that you would like SAMI to recommend potential teammates to you for a school project (e.g., extended class project, final-year team project), please write a paragraph (at least 6 sentences) to introduce yourself to SAMI. 
    
    Consider this a free-flowing essay, where you write different things about you (e.g., where you grew up, your interests and hobbies, your feelings, thoughts, and emotions, your dreams and passions, you career goals, fun facts about you) and let your thoughts flow freely in your writing without pauses. The more you write, the better it will help us during the study!

    \item Below are a number of characteristics that may or may not apply to you. For example, do you agree that you are someone who likes to spend time with others? Please rate each statement on a scale of 1 to 5 (1-Disagree Strongly; 2-Disagree a little; 3-Neither agree nor disagree; 4-Agree a little; 5-Agree strongly) to indicate the extent to which you agree or disagree with that statement. 

    I see myself as someone who... 
    \begin{itemize}
        \item Is talkative
        \item Tends to find fault with others
        \item Does a thorough job
        \item Is depressed, blue
        \item Is original, comes up with new ideas
        \item Is reserved
        \item Is helpful and unselfish with others
        \item Can be somewhat careless
        \item Is relaxed, handles stress well
        \item Is curious about many different things
        \item Is full of energy
        \item Starts quarrels with others
        \item Is a reliable worker
        \item Can be tense
        \item Is ingenious, a deep thinker
        \item Generates a lot of enthusiasm
        \item Has a forgiving nature
        \item Tends to be disorganized
        \item Worries a lot
        \item Has an active imagination
        \item Tends to be quiet
        \item Is generally trusting
        \item Tends to be lazy
        \item Is emotionally stable, not easily upset
        \item Is inventive
        \item Has an assertive personality
        \item Can be cold and aloof
        \item Perseveres until the task is finished
        \item Can be moody
        \item Values artistic, aesthetic experiences
        \item Is sometimes shy, inhibited
        \item Is considerate and kind to almost everyone
        \item Does things efficiently
        \item Remains calm in tense situations
        \item Prefers work that is routine
        \item Is outgoing, sociable
        \item Is sometimes rude to others
        \item Makes plans and follows through with them
        \item Gets nervous easily
        \item Likes to reflect, play with ideas
        \item Has few artistic interests
        \item Likes to cooperate with others
        \item Is easily distracted
        \item Is sophisticated in art, music, or literature
    \end{itemize}
    \item This questionnaire aims at understanding your knowledge and experience with Artificial Intelligence. Please rate each of the following statements on a scale of 1 to 7 (1-Disagree Strongly; 2-Disagree a little; 3-Somewhat Disagree; 4-Neither Agree nor Disagree; 5-Agree a little; 6-Somewhat Agree; 7-Agree Strongly) to indicate the extent to which you agree or disagree with that statement.
    \begin{itemize}
        \item I have knowledge of the types of technology that AI is built on. 
        \item I have knowledge of how AI technology and non-AI technology are distinct.
        \item I have knowledge of use cases for AI technology.
        \item I have knowledge of which human actors beyond programmers are involved to enable human-AI collaboration
        \item I have knowledge of the aspects human actors handle worse than AI.
        \item I have knowledge of the aspects human actors handle better than AI.
        \item I have knowledge of the input data requirements for AI.
        \item I have knowledge of AI processing methods and models.
        \item I have knowledge of using AI output and interpreting it.
        \item I have experience in interaction with different types of AI, like chatbots, visual recognition agents, etc.
        \item I have experience in the usage of AI through frequent interactions in my everyday life
        \item I have experience in designing AI models, for example, a neural network
        \item I have experience in development of AI products
        \item In general, I know the unique facets of AI and humans and their potential roles in human-AI collaboration.
        \item I am knowledgeable about the steps involved in AI decision-making.
        \item Considering all my experience, I am relatively proficient in the field of AI.
    \end{itemize}
    \item How old are you? (Please enter a number)
    \item What is your gender?
    \begin{itemize}
        \item Woman
        \item Man
        \item Non-binary
        \item Prefer not to disclose
        \item Prefer to self-describe: 
    \end{itemize}
    \item What is your current level of study?
    \begin{itemize}
        \item Undergraduate
        \item Master
        \item Doctorate
        \item Other, please specify:
    \end{itemize}
    \item What major(s) are you in?
    \item Roughly how many team projects have you been involved in the past? Please enter a number: 
    \item How would you describe your overall experience with your past team projects?
    \begin{itemize}
        \item Extremely negative
        \item Somewhat negative
        \item Neutral
        \item Somewhat positive
        \item Extremely positive
    \end{itemize}
\end{enumerate}

\section{Study 2 Experiment Survey} \label{supp:s2_exp_survey}
(insert study consent form here)

\noindent \textbf{Baseline Perception Measures}

In this section, you will be shown two samples of a students’ self-introduction paragraph and SAMI’s inferences about the student based on their self-introduction. These samples are only to give you a sense of what SAMI’s inferences look like. You will then be prompted to answer some questions about your perception of SAMI after seeing the samples based on your impression of SAMI. (The two samples shown in this survey are the same as described in Supplementary Materials~\ref{supp:s1_protocol}.)

\begin{enumerate}
    \item In sample \#1:
    \begin{enumerate}
        \item How many of SAMI's inferences about Lin do you believe to be accurate? Please enter a number. 
        \item How many of SAMI's inferences about Lin do you believe to be inaccurate? Please enter a number.
    \end{enumerate}
    \item In sample \#2:
    \begin{enumerate}
        \item How many of SAMI's inferences about Joey do you believe to be accurate? Please enter a number.
        \item How many of SAMI's inferences about Joey do you believe to be inaccurate? Please enter a number. 
    \end{enumerate}
    \item On a scale of 1 to 5, how would you rate the overall accuracy of SAMI’s inferences about students based on the two samples you have seen?
    \begin{itemize}
        \item 1- Not accurate at all
        \item 2
        \item 3- Somewhat accurate
        \item 4
        \item 5- Very accurate
    \end{itemize}
    \item Now that you have seen samples of how SAMI works, please answer the following questionnaire based on your current impression of SAMI. Please rate each of the following statement based on how much you agree with it on a scale of 1 to 5, 1 indicates strongly disagree and 5 indicates strongly agree.
    \begin{itemize}
        \item I believe that there could be negative consequences when using SAMI. 
        \item I feel I must be cautious when using SAMI.
        \item It is risky to interact with SAMI.
        \item I believe that SAMI will act in my best interest.
        \item I believe that SAMI will do its best to help me if I need help.
        \item I believe that SAMI is interested in understanding my needs and preferences.
        \item I think that SAMI is competent and effective in making accurate inferences about me as a person.
        \item I think that SAMI performs its role as a social recommendation agent very well.
        \item I believe that SAMI has all the functionalities I would expect from a social recommendation agent. 
        \item If I use SAMI, i think I would be able to depend on it completely.
        \item I can always rely on SAMI for recommending students that fit my social needs.
        \item I can trust the information presented to me by SAMI. 
    \end{itemize}
    \item Please answer the following questionnaire based on your current impression of SAMI. The following questions will give you a spectrum from one quality to the other on a scale of 1 to 5, such as from “Unkind (1)” to “Kind (5).” Please rate your perception of SAMI along each of these spectrum:
    \begin{figure}[h]
        \centering
        \includegraphics[width=0.5\textwidth]{figures/supp_godspeed_survey.pdf}
        \caption{This figure shows the Godspeed questionnaire we used to measure students' social perceptions of SAMI after seeing the samples.}
        \label{supp:godspeed_survey_s2_base}
    \end{figure}
\end{enumerate}

\noindent \textbf{Experiment Perception Measures}

Thank you! In this section, you will be shown your own self-introduction paragraph and SAMI’s inferences about you based on your self-introduction. 

(Showed participant's self-intro and SAMI's fabricated inferences about the participant).

\begin{enumerate}
    \item In SAMI's inferences about you:
    \begin{enumerate}
        \item How many of SAMI's inferences about you do you think are accurate? Please enter a number.
        \item How many of SAMI's inferences about you do you think are inaccurate? Please enter a number. 
    \end{enumerate}
    \item On a scale of 1 to 5, how would you rate the overall accuracy of SAMI’s inferences?
    \begin{itemize}
        \item 1- Not accurate at all
        \item 2
        \item 3- Somewhat accurate
        \item 4
        \item 5- Very accurate
    \end{itemize}
    \item Now that you have seen samples of how SAMI works, please answer the following questionnaire based on your current impression of SAMI. Please rate each of the following statement based on how much you agree with it on a scale of 1 to 5, 1 indicates strongly disagree and 5 indicates strongly agree.
    \begin{itemize}
        \item I believe that there could be negative consequences when using SAMI. 
        \item I feel I must be cautious when using SAMI.
        \item It is risky to interact with SAMI.
        \item I believe that SAMI will act in my best interest.
        \item I believe that SAMI will do its best to help me if I need help.
        \item I believe that SAMI is interested in understanding my needs and preferences.
        \item I think that SAMI is competent and effective in making accurate inferences about me as a person.
        \item I think that SAMI performs its role as a social recommendation agent very well.
        \item I believe that SAMI has all the functionalities I would expect from a social recommendation agent. 
        \item If I use SAMI, i think I would be able to depend on it completely.
        \item I can always rely on SAMI for recommending students that fit my social needs.
        \item I can trust the information presented to me by SAMI. 
    \end{itemize}
    \item Please answer the following questionnaire based on your current impression of SAMI. The following questions will give you a spectrum from one quality to the other on a scale of 1 to 5, such as from “Unkind (1)” to “Kind (5).” Please rate your perception of SAMI along each of these spectrum:
    \begin{figure}[h]
        \centering
        \includegraphics[width=0.5\textwidth]{figures/supp_godspeed_survey.pdf}
        \caption{This figure shows the Godspeed questionnaire we used to measure students' social perceptions of SAMI after seeing the samples.}
        \label{supp:godspeed_survey_s2_exp}
    \end{figure}
\end{enumerate}

[Insert debriefing form here]

%% file: 0main.bbl

\begin{thebibliography}{63}


\ifx \showCODEN    \undefined \def \showCODEN     #1{\unskip}     \fi
\ifx \showDOI      \undefined \def \showDOI       #1{#1}\fi
\ifx \showISBNx    \undefined \def \showISBNx     #1{\unskip}     \fi
\ifx \showISBNxiii \undefined \def \showISBNxiii  #1{\unskip}     \fi
\ifx \showISSN     \undefined \def \showISSN      #1{\unskip}     \fi
\ifx \showLCCN     \undefined \def \showLCCN      #1{\unskip}     \fi
\ifx \shownote     \undefined \def \shownote      #1{#1}          \fi
\ifx \showarticletitle \undefined \def \showarticletitle #1{#1}   \fi
\ifx \showURL      \undefined \def \showURL       {\relax}        \fi
\providecommand\bibfield[2]{#2}
\providecommand\bibinfo[2]{#2}
\providecommand\natexlab[1]{#1}
\providecommand\showeprint[2][]{arXiv:#2}

\bibitem[Alberola et~al\mbox{.}(2016)]%
        {alberola2016artificial}
\bibfield{author}{\bibinfo{person}{Juan~M Alberola}, \bibinfo{person}{Elena Del~Val}, \bibinfo{person}{Victor Sanchez-Anguix}, \bibinfo{person}{Alberto Palomares}, {and} \bibinfo{person}{Maria~Dolores Teruel}.} \bibinfo{year}{2016}\natexlab{}.
\newblock \showarticletitle{An artificial intelligence tool for heterogeneous team formation in the classroom}.
\newblock \bibinfo{journal}{\emph{Knowledge-Based Systems}}  \bibinfo{volume}{101} (\bibinfo{year}{2016}), \bibinfo{pages}{1--14}.
\newblock


\bibitem[Ashktorab et~al\mbox{.}(2019)]%
        {ashktorab2019resilient}
\bibfield{author}{\bibinfo{person}{Zahra Ashktorab}, \bibinfo{person}{Mohit Jain}, \bibinfo{person}{Q~Vera Liao}, {and} \bibinfo{person}{Justin~D Weisz}.} \bibinfo{year}{2019}\natexlab{}.
\newblock \showarticletitle{Resilient chatbots: Repair strategy preferences for conversational breakdowns}. In \bibinfo{booktitle}{\emph{Proceedings of the 2019 CHI conference on human factors in computing systems}}. \bibinfo{pages}{1--12}.
\newblock


\bibitem[Banas et~al\mbox{.}(2022)]%
        {banas2022machine}
\bibfield{author}{\bibinfo{person}{John~A Banas}, \bibinfo{person}{Nicholas~A Palomares}, \bibinfo{person}{Adam~S Richards}, \bibinfo{person}{David~M Keating}, \bibinfo{person}{Nick Joyce}, {and} \bibinfo{person}{Stephen~A Rains}.} \bibinfo{year}{2022}\natexlab{}.
\newblock \showarticletitle{When machine and bandwagon heuristics compete: Understanding users’ response to conflicting AI and crowdsourced fact-checking}.
\newblock \bibinfo{journal}{\emph{Human Communication Research}} \bibinfo{volume}{48}, \bibinfo{number}{3} (\bibinfo{year}{2022}), \bibinfo{pages}{430--461}.
\newblock


\bibitem[Bartneck et~al\mbox{.}(2009)]%
        {bartneck2009measurement}
\bibfield{author}{\bibinfo{person}{Christoph Bartneck}, \bibinfo{person}{Dana Kuli{\'c}}, \bibinfo{person}{Elizabeth Croft}, {and} \bibinfo{person}{Susana Zoghbi}.} \bibinfo{year}{2009}\natexlab{}.
\newblock \showarticletitle{Measurement instruments for the anthropomorphism, animacy, likeability, perceived intelligence, and perceived safety of robots}.
\newblock \bibinfo{journal}{\emph{International journal of social robotics}}  \bibinfo{volume}{1} (\bibinfo{year}{2009}), \bibinfo{pages}{71--81}.
\newblock


\bibitem[Braun and Clarke(2006)]%
        {braun2006using}
\bibfield{author}{\bibinfo{person}{Virginia Braun} {and} \bibinfo{person}{Victoria Clarke}.} \bibinfo{year}{2006}\natexlab{}.
\newblock \showarticletitle{Using thematic analysis in psychology}.
\newblock \bibinfo{journal}{\emph{Qualitative research in psychology}} \bibinfo{volume}{3}, \bibinfo{number}{2} (\bibinfo{year}{2006}), \bibinfo{pages}{77--101}.
\newblock


\bibitem[Braun and Clarke(2019)]%
        {braun2019reflecting}
\bibfield{author}{\bibinfo{person}{Virginia Braun} {and} \bibinfo{person}{Victoria Clarke}.} \bibinfo{year}{2019}\natexlab{}.
\newblock \showarticletitle{Reflecting on reflexive thematic analysis}.
\newblock \bibinfo{journal}{\emph{Qualitative research in sport, exercise and health}} \bibinfo{volume}{11}, \bibinfo{number}{4} (\bibinfo{year}{2019}), \bibinfo{pages}{589--597}.
\newblock


\bibitem[B{\"u}hrke et~al\mbox{.}(2021)]%
        {buhrke2021making}
\bibfield{author}{\bibinfo{person}{Johannes B{\"u}hrke}, \bibinfo{person}{Alfred~Benedikt Brendel}, \bibinfo{person}{Sascha Lichtenberg}, \bibinfo{person}{Maike Greve}, {and} \bibinfo{person}{Milad Mirbabaie}.} \bibinfo{year}{2021}\natexlab{}.
\newblock \showarticletitle{Is making mistakes human? On the perception of typing errors in chatbot communication}.
\newblock  (\bibinfo{year}{2021}).
\newblock


\bibitem[Chen et~al\mbox{.}(2023a)]%
        {chen2023machine}
\bibfield{author}{\bibinfo{person}{Chacha Chen}, \bibinfo{person}{Shi Feng}, \bibinfo{person}{Amit Sharma}, {and} \bibinfo{person}{Chenhao Tan}.} \bibinfo{year}{2023}\natexlab{a}.
\newblock \showarticletitle{Machine Explanations and Human Understanding}. In \bibinfo{booktitle}{\emph{Proceedings of the 2023 ACM Conference on Fairness, Accountability, and Transparency}}. \bibinfo{pages}{1--1}.
\newblock


\bibitem[Chen et~al\mbox{.}(2023b)]%
        {chen2023understanding}
\bibfield{author}{\bibinfo{person}{Valerie Chen}, \bibinfo{person}{Q~Vera Liao}, \bibinfo{person}{Jennifer~Wortman Vaughan}, {and} \bibinfo{person}{Gagan Bansal}.} \bibinfo{year}{2023}\natexlab{b}.
\newblock \showarticletitle{Understanding the role of human intuition on reliance in human-AI decision-making with explanations}.
\newblock \bibinfo{journal}{\emph{arXiv preprint arXiv:2301.07255}} (\bibinfo{year}{2023}).
\newblock


\bibitem[Chi et~al\mbox{.}(1994)]%
        {chi1994things}
\bibfield{author}{\bibinfo{person}{Michelene~TH Chi}, \bibinfo{person}{James~D Slotta}, {and} \bibinfo{person}{Nicholas De~Leeuw}.} \bibinfo{year}{1994}\natexlab{}.
\newblock \showarticletitle{From things to processes: A theory of conceptual change for learning science concepts}.
\newblock \bibinfo{journal}{\emph{Learning and instruction}} \bibinfo{volume}{4}, \bibinfo{number}{1} (\bibinfo{year}{1994}), \bibinfo{pages}{27--43}.
\newblock


\bibitem[Clausen et~al\mbox{.}(2023)]%
        {clausen2023exploring}
\bibfield{author}{\bibinfo{person}{Mikkel Clausen}, \bibinfo{person}{Mikkel~Peter Kyhn}, \bibinfo{person}{Eleftherios Papachristos}, {and} \bibinfo{person}{Timothy Merritt}.} \bibinfo{year}{2023}\natexlab{}.
\newblock \showarticletitle{Exploring Humor as a Repair Strategy During Communication Breakdowns with Voice Assistants}. In \bibinfo{booktitle}{\emph{Proceedings of the 5th International Conference on Conversational User Interfaces}}. \bibinfo{pages}{1--9}.
\newblock


\bibitem[de~S{\'a}~Siqueira et~al\mbox{.}(2023)]%
        {de2023we}
\bibfield{author}{\bibinfo{person}{Marianna~A de S{\'a}~Siqueira}, \bibinfo{person}{Barbara~CN M{\"u}ller}, {and} \bibinfo{person}{Tibor Bosse}.} \bibinfo{year}{2023}\natexlab{}.
\newblock \showarticletitle{When do we accept mistakes from chatbots? The impact of human-like communication on user experience in chatbots that make mistakes}.
\newblock \bibinfo{journal}{\emph{International Journal of Human--Computer Interaction}} (\bibinfo{year}{2023}), \bibinfo{pages}{1--11}.
\newblock


\bibitem[DeVito et~al\mbox{.}(2017)]%
        {devito2017platforms}
\bibfield{author}{\bibinfo{person}{Michael~A DeVito}, \bibinfo{person}{Jeremy Birnholtz}, {and} \bibinfo{person}{Jeffery~T Hancock}.} \bibinfo{year}{2017}\natexlab{}.
\newblock \showarticletitle{Platforms, people, and perception: Using affordances to understand self-presentation on social media}. In \bibinfo{booktitle}{\emph{Proceedings of the 2017 ACM conference on computer supported cooperative work and social computing}}. \bibinfo{pages}{740--754}.
\newblock


\bibitem[Dickson and Kelly(1985)]%
        {dickson1985barnum}
\bibfield{author}{\bibinfo{person}{DH Dickson} {and} \bibinfo{person}{IW Kelly}.} \bibinfo{year}{1985}\natexlab{}.
\newblock \showarticletitle{The ‘Barnum Effect’in personality assessment: A review of the literature}.
\newblock \bibinfo{journal}{\emph{Psychological reports}} \bibinfo{volume}{57}, \bibinfo{number}{2} (\bibinfo{year}{1985}), \bibinfo{pages}{367--382}.
\newblock


\bibitem[DiSalvo et~al\mbox{.}(2022)]%
        {disalvo2022reading}
\bibfield{author}{\bibinfo{person}{Betsy DiSalvo}, \bibinfo{person}{Dheeraj Bandaru}, \bibinfo{person}{Qiaosi Wang}, \bibinfo{person}{Hong Li}, {and} \bibinfo{person}{Thomas Pl{\"o}tz}.} \bibinfo{year}{2022}\natexlab{}.
\newblock \showarticletitle{Reading the Room: Automated, Momentary Assessment of Student Engagement in the Classroom: Are We There Yet?}
\newblock \bibinfo{journal}{\emph{Proceedings of the ACM on Interactive, Mobile, Wearable and Ubiquitous Technologies}} \bibinfo{volume}{6}, \bibinfo{number}{3} (\bibinfo{year}{2022}), \bibinfo{pages}{1--26}.
\newblock


\bibitem[Druga et~al\mbox{.}(2017)]%
        {druga2017hey}
\bibfield{author}{\bibinfo{person}{Stefania Druga}, \bibinfo{person}{Randi Williams}, \bibinfo{person}{Cynthia Breazeal}, {and} \bibinfo{person}{Mitchel Resnick}.} \bibinfo{year}{2017}\natexlab{}.
\newblock \showarticletitle{" Hey Google is it ok if I eat you?" Initial explorations in child-agent interaction}. In \bibinfo{booktitle}{\emph{Proceedings of the 2017 conference on interaction design and children}}. \bibinfo{pages}{595--600}.
\newblock


\bibitem[Edwards et~al\mbox{.}(2016)]%
        {edwards2016robots}
\bibfield{author}{\bibinfo{person}{Autumn Edwards}, \bibinfo{person}{Chad Edwards}, \bibinfo{person}{Patric~R Spence}, \bibinfo{person}{Christina Harris}, {and} \bibinfo{person}{Andrew Gambino}.} \bibinfo{year}{2016}\natexlab{}.
\newblock \showarticletitle{Robots in the classroom: Differences in students’ perceptions of credibility and learning between “teacher as robot” and “robot as teacher”}.
\newblock \bibinfo{journal}{\emph{Computers in Human Behavior}}  \bibinfo{volume}{65} (\bibinfo{year}{2016}), \bibinfo{pages}{627--634}.
\newblock


\bibitem[Ehsan and Riedl(2020)]%
        {ehsan2020human}
\bibfield{author}{\bibinfo{person}{Upol Ehsan} {and} \bibinfo{person}{Mark~O Riedl}.} \bibinfo{year}{2020}\natexlab{}.
\newblock \showarticletitle{Human-centered explainable ai: Towards a reflective sociotechnical approach}. In \bibinfo{booktitle}{\emph{HCI International 2020-Late Breaking Papers: Multimodality and Intelligence: 22nd HCI International Conference, HCII 2020, Copenhagen, Denmark, July 19--24, 2020, Proceedings 22}}. Springer, \bibinfo{pages}{449--466}.
\newblock


\bibitem[Eslami et~al\mbox{.}(2016)]%
        {eslami2016first}
\bibfield{author}{\bibinfo{person}{Motahhare Eslami}, \bibinfo{person}{Karrie Karahalios}, \bibinfo{person}{Christian Sandvig}, \bibinfo{person}{Kristen Vaccaro}, \bibinfo{person}{Aimee Rickman}, \bibinfo{person}{Kevin Hamilton}, {and} \bibinfo{person}{Alex Kirlik}.} \bibinfo{year}{2016}\natexlab{}.
\newblock \showarticletitle{First I" like" it, then I hide it: Folk Theories of Social Feeds}. In \bibinfo{booktitle}{\emph{Proceedings of the 2016 cHI conference on human factors in computing systems}}. \bibinfo{pages}{2371--2382}.
\newblock


\bibitem[Eslami et~al\mbox{.}(2019)]%
        {eslami2019user}
\bibfield{author}{\bibinfo{person}{Motahhare Eslami}, \bibinfo{person}{Kristen Vaccaro}, \bibinfo{person}{Min~Kyung Lee}, \bibinfo{person}{Amit Elazari Bar~On}, \bibinfo{person}{Eric Gilbert}, {and} \bibinfo{person}{Karrie Karahalios}.} \bibinfo{year}{2019}\natexlab{}.
\newblock \showarticletitle{User attitudes towards algorithmic opacity and transparency in online reviewing platforms}. In \bibinfo{booktitle}{\emph{Proceedings of the 2019 CHI Conference on Human Factors in Computing Systems}}. \bibinfo{pages}{1--14}.
\newblock


\bibitem[Gero et~al\mbox{.}(2020)]%
        {gero2020mental}
\bibfield{author}{\bibinfo{person}{Katy~Ilonka Gero}, \bibinfo{person}{Zahra Ashktorab}, \bibinfo{person}{Casey Dugan}, \bibinfo{person}{Qian Pan}, \bibinfo{person}{James Johnson}, \bibinfo{person}{Werner Geyer}, \bibinfo{person}{Maria Ruiz}, \bibinfo{person}{Sarah Miller}, \bibinfo{person}{David~R Millen}, \bibinfo{person}{Murray Campbell}, {et~al\mbox{.}}} \bibinfo{year}{2020}\natexlab{}.
\newblock \showarticletitle{Mental models of AI agents in a cooperative game setting}. In \bibinfo{booktitle}{\emph{Proceedings of the 2020 chi conference on human factors in computing systems}}. \bibinfo{pages}{1--12}.
\newblock


\bibitem[Gompei and Umemuro(2015)]%
        {gompei2015robot}
\bibfield{author}{\bibinfo{person}{Takayuki Gompei} {and} \bibinfo{person}{Hiroyuki Umemuro}.} \bibinfo{year}{2015}\natexlab{}.
\newblock \showarticletitle{A robot's slip of the tongue: Effect of speech error on the familiarity of a humanoid robot}. In \bibinfo{booktitle}{\emph{2015 24th IEEE international symposium on robot and human interactive communication (RO-MAN)}}. IEEE, \bibinfo{pages}{331--336}.
\newblock


\bibitem[Gou et~al\mbox{.}(2014)]%
        {gou2014knowme}
\bibfield{author}{\bibinfo{person}{Liang Gou}, \bibinfo{person}{Michelle~X Zhou}, {and} \bibinfo{person}{Huahai Yang}.} \bibinfo{year}{2014}\natexlab{}.
\newblock \showarticletitle{KnowMe and ShareMe: understanding automatically discovered personality traits from social media and user sharing preferences}. In \bibinfo{booktitle}{\emph{Proceedings of the SIGCHI conference on human factors in computing systems}}. \bibinfo{pages}{955--964}.
\newblock


\bibitem[Gulati et~al\mbox{.}(2019)]%
        {gulati2019design}
\bibfield{author}{\bibinfo{person}{Siddharth Gulati}, \bibinfo{person}{Sonia Sousa}, {and} \bibinfo{person}{David Lamas}.} \bibinfo{year}{2019}\natexlab{}.
\newblock \showarticletitle{Design, development and evaluation of a human-computer trust scale}.
\newblock \bibinfo{journal}{\emph{Behaviour \& Information Technology}} \bibinfo{volume}{38}, \bibinfo{number}{10} (\bibinfo{year}{2019}), \bibinfo{pages}{1004--1015}.
\newblock


\bibitem[Guzman(2020)]%
        {guzman2020ontological}
\bibfield{author}{\bibinfo{person}{Andrea~L Guzman}.} \bibinfo{year}{2020}\natexlab{}.
\newblock \showarticletitle{Ontological boundaries between humans and computers and the implications for human-machine communication}.
\newblock \bibinfo{journal}{\emph{Human-Machine Communication}}  \bibinfo{volume}{1} (\bibinfo{year}{2020}), \bibinfo{pages}{37--54}.
\newblock


\bibitem[Hall and Caton(2017)]%
        {hall2017say}
\bibfield{author}{\bibinfo{person}{Margeret Hall} {and} \bibinfo{person}{Simon Caton}.} \bibinfo{year}{2017}\natexlab{}.
\newblock \showarticletitle{Am I who I say I am? Unobtrusive self-representation and personality recognition on Facebook}.
\newblock \bibinfo{journal}{\emph{PloS one}} \bibinfo{volume}{12}, \bibinfo{number}{9} (\bibinfo{year}{2017}), \bibinfo{pages}{e0184417}.
\newblock


\bibitem[Hollis et~al\mbox{.}(2018)]%
        {hollis2018being}
\bibfield{author}{\bibinfo{person}{Victoria Hollis}, \bibinfo{person}{Alon Pekurovsky}, \bibinfo{person}{Eunika Wu}, {and} \bibinfo{person}{Steve Whittaker}.} \bibinfo{year}{2018}\natexlab{}.
\newblock \showarticletitle{On being told how we feel: how algorithmic sensor feedback influences emotion perception}.
\newblock \bibinfo{journal}{\emph{Proceedings of the ACM on Interactive, Mobile, Wearable and Ubiquitous Technologies}} \bibinfo{volume}{2}, \bibinfo{number}{3} (\bibinfo{year}{2018}), \bibinfo{pages}{1--31}.
\newblock


\bibitem[Honig and Oron-Gilad(2018)]%
        {honig2018understanding}
\bibfield{author}{\bibinfo{person}{Shanee Honig} {and} \bibinfo{person}{Tal Oron-Gilad}.} \bibinfo{year}{2018}\natexlab{}.
\newblock \showarticletitle{Understanding and resolving failures in human-robot interaction: Literature review and model development}.
\newblock \bibinfo{journal}{\emph{Frontiers in psychology}}  \bibinfo{volume}{9} (\bibinfo{year}{2018}), \bibinfo{pages}{861}.
\newblock


\bibitem[Hsu et~al\mbox{.}(2021)]%
        {hsu2021attitudes}
\bibfield{author}{\bibinfo{person}{Silas Hsu}, \bibinfo{person}{Tiffany~Wenting Li}, \bibinfo{person}{Zhilin Zhang}, \bibinfo{person}{Max Fowler}, \bibinfo{person}{Craig Zilles}, {and} \bibinfo{person}{Karrie Karahalios}.} \bibinfo{year}{2021}\natexlab{}.
\newblock \showarticletitle{Attitudes surrounding an imperfect AI autograder}. In \bibinfo{booktitle}{\emph{Proceedings of the 2021 CHI conference on human factors in computing systems}}. \bibinfo{pages}{1--15}.
\newblock


\bibitem[Jahanbakhsh et~al\mbox{.}(2017)]%
        {jahanbakhsh2017you}
\bibfield{author}{\bibinfo{person}{Farnaz Jahanbakhsh}, \bibinfo{person}{Wai-Tat Fu}, \bibinfo{person}{Karrie Karahalios}, \bibinfo{person}{Darko Marinov}, {and} \bibinfo{person}{Brian Bailey}.} \bibinfo{year}{2017}\natexlab{}.
\newblock \showarticletitle{You want me to work with who? Stakeholder perceptions of automated team formation in project-based courses}. In \bibinfo{booktitle}{\emph{Proceedings of the 2017 CHI conference on human factors in computing systems}}. \bibinfo{pages}{3201--3212}.
\newblock


\bibitem[Kapania et~al\mbox{.}(2022)]%
        {kapania2022because}
\bibfield{author}{\bibinfo{person}{Shivani Kapania}, \bibinfo{person}{Oliver Siy}, \bibinfo{person}{Gabe Clapper}, \bibinfo{person}{Azhagu~Meena SP}, {and} \bibinfo{person}{Nithya Sambasivan}.} \bibinfo{year}{2022}\natexlab{}.
\newblock \showarticletitle{” Because AI is 100\% right and safe”: User attitudes and sources of AI authority in India}. In \bibinfo{booktitle}{\emph{Proceedings of the 2022 CHI Conference on Human Factors in Computing Systems}}. \bibinfo{pages}{1--18}.
\newblock


\bibitem[Kim et~al\mbox{.}(2020)]%
        {kim2020understanding}
\bibfield{author}{\bibinfo{person}{Seoyoung Kim}, \bibinfo{person}{Arti Thakur}, {and} \bibinfo{person}{Juho Kim}.} \bibinfo{year}{2020}\natexlab{}.
\newblock \showarticletitle{Understanding Users' Perception Towards Automated Personality Detection with Group-specific Behavioral Data}. In \bibinfo{booktitle}{\emph{Proceedings of the 2020 CHI Conference on Human Factors in Computing Systems}}. \bibinfo{pages}{1--12}.
\newblock


\bibitem[Kocielnik et~al\mbox{.}(2019)]%
        {kocielnik2019will}
\bibfield{author}{\bibinfo{person}{Rafal Kocielnik}, \bibinfo{person}{Saleema Amershi}, {and} \bibinfo{person}{Paul~N Bennett}.} \bibinfo{year}{2019}\natexlab{}.
\newblock \showarticletitle{Will you accept an imperfect ai? exploring designs for adjusting end-user expectations of ai systems}. In \bibinfo{booktitle}{\emph{Proceedings of the 2019 CHI Conference on Human Factors in Computing Systems}}. \bibinfo{pages}{1--14}.
\newblock


\bibitem[Law et~al\mbox{.}(2017)]%
        {law2017wizard}
\bibfield{author}{\bibinfo{person}{Edith Law}, \bibinfo{person}{Vicky Cai}, \bibinfo{person}{Qi~Feng Liu}, \bibinfo{person}{Sajin Sasy}, \bibinfo{person}{Joslin Goh}, \bibinfo{person}{Alex Blidaru}, {and} \bibinfo{person}{Dana Kuli{\'c}}.} \bibinfo{year}{2017}\natexlab{}.
\newblock \showarticletitle{A wizard-of-oz study of curiosity in human-robot interaction}. In \bibinfo{booktitle}{\emph{2017 26th IEEE International Symposium on Robot and Human Interactive Communication (RO-MAN)}}. IEEE, \bibinfo{pages}{607--614}.
\newblock


\bibitem[Lee et~al\mbox{.}(2010)]%
        {lee2010gracefully}
\bibfield{author}{\bibinfo{person}{Min~Kyung Lee}, \bibinfo{person}{Sara Kiesler}, \bibinfo{person}{Jodi Forlizzi}, \bibinfo{person}{Siddhartha Srinivasa}, {and} \bibinfo{person}{Paul Rybski}.} \bibinfo{year}{2010}\natexlab{}.
\newblock \showarticletitle{Gracefully mitigating breakdowns in robotic services}. In \bibinfo{booktitle}{\emph{2010 5th ACM/IEEE International Conference on Human-Robot Interaction (HRI)}}. IEEE, \bibinfo{pages}{203--210}.
\newblock


\bibitem[Liao and Varshney(2021)]%
        {liao2021human}
\bibfield{author}{\bibinfo{person}{Q~Vera Liao} {and} \bibinfo{person}{Kush~R Varshney}.} \bibinfo{year}{2021}\natexlab{}.
\newblock \showarticletitle{Human-centered explainable ai (xai): From algorithms to user experiences}.
\newblock \bibinfo{journal}{\emph{arXiv preprint arXiv:2110.10790}} (\bibinfo{year}{2021}).
\newblock


\bibitem[Liao and Tyson(2021)]%
        {liao2021crystal}
\bibfield{author}{\bibinfo{person}{Tony Liao} {and} \bibinfo{person}{Olivia Tyson}.} \bibinfo{year}{2021}\natexlab{}.
\newblock \showarticletitle{“Crystal Is Creepy, but Cool”: Mapping Folk Theories and Responses to Automated Personality Recognition Algorithms}.
\newblock \bibinfo{journal}{\emph{Social Media+ Society}} \bibinfo{volume}{7}, \bibinfo{number}{2} (\bibinfo{year}{2021}), \bibinfo{pages}{20563051211010170}.
\newblock


\bibitem[Long and Magerko(2020)]%
        {long2020ai}
\bibfield{author}{\bibinfo{person}{Duri Long} {and} \bibinfo{person}{Brian Magerko}.} \bibinfo{year}{2020}\natexlab{}.
\newblock \showarticletitle{What is AI literacy? Competencies and design considerations}. In \bibinfo{booktitle}{\emph{Proceedings of the 2020 CHI conference on human factors in computing systems}}. \bibinfo{pages}{1--16}.
\newblock


\bibitem[Lykourentzou et~al\mbox{.}(2016)]%
        {lykourentzou2016personality}
\bibfield{author}{\bibinfo{person}{Ioanna Lykourentzou}, \bibinfo{person}{Angeliki Antoniou}, \bibinfo{person}{Yannick Naudet}, {and} \bibinfo{person}{Steven~P Dow}.} \bibinfo{year}{2016}\natexlab{}.
\newblock \showarticletitle{Personality matters: Balancing for personality types leads to better outcomes for crowd teams}. In \bibinfo{booktitle}{\emph{Proceedings of the 19th ACM Conference on Computer-Supported Cooperative Work \& Social Computing}}. \bibinfo{pages}{260--273}.
\newblock


\bibitem[Mahmood et~al\mbox{.}(2022)]%
        {mahmood2022owning}
\bibfield{author}{\bibinfo{person}{Amama Mahmood}, \bibinfo{person}{Jeanie~W Fung}, \bibinfo{person}{Isabel Won}, {and} \bibinfo{person}{Chien-Ming Huang}.} \bibinfo{year}{2022}\natexlab{}.
\newblock \showarticletitle{Owning mistakes sincerely: Strategies for mitigating AI errors}. In \bibinfo{booktitle}{\emph{Proceedings of the 2022 CHI Conference on Human Factors in Computing Systems}}. \bibinfo{pages}{1--11}.
\newblock


\bibitem[Miller(2019)]%
        {miller2019explanation}
\bibfield{author}{\bibinfo{person}{Tim Miller}.} \bibinfo{year}{2019}\natexlab{}.
\newblock \showarticletitle{Explanation in artificial intelligence: Insights from the social sciences}.
\newblock \bibinfo{journal}{\emph{Artificial intelligence}}  \bibinfo{volume}{267} (\bibinfo{year}{2019}), \bibinfo{pages}{1--38}.
\newblock


\bibitem[Mirnig et~al\mbox{.}(2017)]%
        {mirnig2017err}
\bibfield{author}{\bibinfo{person}{Nicole Mirnig}, \bibinfo{person}{Gerald Stollnberger}, \bibinfo{person}{Markus Miksch}, \bibinfo{person}{Susanne Stadler}, \bibinfo{person}{Manuel Giuliani}, {and} \bibinfo{person}{Manfred Tscheligi}.} \bibinfo{year}{2017}\natexlab{}.
\newblock \showarticletitle{To err is robot: How humans assess and act toward an erroneous social robot}.
\newblock \bibinfo{journal}{\emph{Frontiers in Robotics and AI}}  \bibinfo{volume}{4} (\bibinfo{year}{2017}), \bibinfo{pages}{21}.
\newblock


\bibitem[Nass et~al\mbox{.}(1994)]%
        {nass1994computers}
\bibfield{author}{\bibinfo{person}{Clifford Nass}, \bibinfo{person}{Jonathan Steuer}, {and} \bibinfo{person}{Ellen~R Tauber}.} \bibinfo{year}{1994}\natexlab{}.
\newblock \showarticletitle{Computers are social actors}. In \bibinfo{booktitle}{\emph{Proceedings of the SIGCHI conference on Human factors in computing systems}}. \bibinfo{pages}{72--78}.
\newblock


\bibitem[Oh and Park(2022)]%
        {oh2022you}
\bibfield{author}{\bibinfo{person}{Soyoung Oh} {and} \bibinfo{person}{Eunil Park}.} \bibinfo{year}{2022}\natexlab{}.
\newblock \showarticletitle{Are you aware of what you are watching? Role of machine heuristic in online content recommendations}.
\newblock \bibinfo{journal}{\emph{arXiv preprint arXiv:2203.08373}} (\bibinfo{year}{2022}).
\newblock


\bibitem[{\"O}zdemir and Clark(2007)]%
        {ozdemir2007overview}
\bibfield{author}{\bibinfo{person}{G{\"o}khan {\"O}zdemir} {and} \bibinfo{person}{Douglas~B Clark}.} \bibinfo{year}{2007}\natexlab{}.
\newblock \showarticletitle{An overview of conceptual change theories}.
\newblock \bibinfo{journal}{\emph{Eurasia Journal of Mathematics, Science and Technology Education}} \bibinfo{volume}{3}, \bibinfo{number}{4} (\bibinfo{year}{2007}), \bibinfo{pages}{351--361}.
\newblock


\bibitem[Pinski and Benlian(2023)]%
        {pinski2023ai}
\bibfield{author}{\bibinfo{person}{Marc Pinski} {and} \bibinfo{person}{Alexander Benlian}.} \bibinfo{year}{2023}\natexlab{}.
\newblock \showarticletitle{AI Literacy-Towards Measuring Human Competency in Artificial Intelligence}.
\newblock  (\bibinfo{year}{2023}).
\newblock


\bibitem[Rao et~al\mbox{.}(2015)]%
        {rao2015they}
\bibfield{author}{\bibinfo{person}{Ashwini Rao}, \bibinfo{person}{Florian Schaub}, {and} \bibinfo{person}{Norman Sadeh}.} \bibinfo{year}{2015}\natexlab{}.
\newblock \showarticletitle{What do they know about me? Contents and concerns of online behavioral profiles}.
\newblock \bibinfo{journal}{\emph{arXiv preprint arXiv:1506.01675}} (\bibinfo{year}{2015}).
\newblock


\bibitem[Roesler et~al\mbox{.}(2020)]%
        {roesler2020effect}
\bibfield{author}{\bibinfo{person}{Eileen Roesler}, \bibinfo{person}{Linda Onnasch}, {and} \bibinfo{person}{Julia~I Majer}.} \bibinfo{year}{2020}\natexlab{}.
\newblock \showarticletitle{The effect of anthropomorphism and failure comprehensibility on human-robot trust}. In \bibinfo{booktitle}{\emph{Proceedings of the human factors and ergonomics society annual meeting}}, Vol.~\bibinfo{volume}{64}. SAGE Publications Sage CA: Los Angeles, CA, \bibinfo{pages}{107--111}.
\newblock


\bibitem[Rossi et~al\mbox{.}(2017)]%
        {rossi2017timing}
\bibfield{author}{\bibinfo{person}{Alessandra Rossi}, \bibinfo{person}{Kerstin Dautenhahn}, \bibinfo{person}{Kheng~Lee Koay}, {and} \bibinfo{person}{Michael~L Walters}.} \bibinfo{year}{2017}\natexlab{}.
\newblock \showarticletitle{How the timing and magnitude of robot errors influence peoples’ trust of robots in an emergency scenario}. In \bibinfo{booktitle}{\emph{Social Robotics: 9th International Conference, ICSR 2017, Tsukuba, Japan, November 22-24, 2017, Proceedings 9}}. Springer, \bibinfo{pages}{42--52}.
\newblock


\bibitem[Salem et~al\mbox{.}(2015)]%
        {salem2015would}
\bibfield{author}{\bibinfo{person}{Maha Salem}, \bibinfo{person}{Gabriella Lakatos}, \bibinfo{person}{Farshid Amirabdollahian}, {and} \bibinfo{person}{Kerstin Dautenhahn}.} \bibinfo{year}{2015}\natexlab{}.
\newblock \showarticletitle{Would you trust a (faulty) robot? Effects of error, task type and personality on human-robot cooperation and trust}. In \bibinfo{booktitle}{\emph{Proceedings of the tenth annual ACM/IEEE international conference on human-robot interaction}}. \bibinfo{pages}{141--148}.
\newblock


\bibitem[Sarkar et~al\mbox{.}(2017)]%
        {sarkar2017effects}
\bibfield{author}{\bibinfo{person}{Satragni Sarkar}, \bibinfo{person}{Dejanira Araiza-Illan}, {and} \bibinfo{person}{Kerstin Eder}.} \bibinfo{year}{2017}\natexlab{}.
\newblock \showarticletitle{Effects of faults, experience, and personality on trust in a robot co-worker}.
\newblock \bibinfo{journal}{\emph{arXiv preprint arXiv:1703.02335}} (\bibinfo{year}{2017}).
\newblock


\bibitem[Schoeffer et~al\mbox{.}(2022)]%
        {schoeffer2022there}
\bibfield{author}{\bibinfo{person}{Jakob Schoeffer}, \bibinfo{person}{Niklas Kuehl}, {and} \bibinfo{person}{Yvette Machowski}.} \bibinfo{year}{2022}\natexlab{}.
\newblock \showarticletitle{“There is not enough information”: On the effects of explanations on perceptions of informational fairness and trustworthiness in automated decision-making}. In \bibinfo{booktitle}{\emph{Proceedings of the 2022 ACM Conference on Fairness, Accountability, and Transparency}}. \bibinfo{pages}{1616--1628}.
\newblock


\bibitem[Shin(2022)]%
        {shin2022people}
\bibfield{author}{\bibinfo{person}{Donghee Shin}.} \bibinfo{year}{2022}\natexlab{}.
\newblock \showarticletitle{How do people judge the credibility of algorithmic sources?}
\newblock \bibinfo{journal}{\emph{Ai \& Society}} (\bibinfo{year}{2022}), \bibinfo{pages}{1--16}.
\newblock


\bibitem[Slotta(2011)]%
        {slotta2011defense}
\bibfield{author}{\bibinfo{person}{James~D Slotta}.} \bibinfo{year}{2011}\natexlab{}.
\newblock \showarticletitle{In defense of Chi's ontological incompatibility hypothesis}.
\newblock \bibinfo{journal}{\emph{The Journal of the Learning Sciences}} \bibinfo{volume}{20}, \bibinfo{number}{1} (\bibinfo{year}{2011}), \bibinfo{pages}{151--162}.
\newblock


\bibitem[Sundar(2020)]%
        {sundar2020rise}
\bibfield{author}{\bibinfo{person}{S~Shyam Sundar}.} \bibinfo{year}{2020}\natexlab{}.
\newblock \showarticletitle{Rise of machine agency: A framework for studying the psychology of human--AI interaction (HAII)}.
\newblock \bibinfo{journal}{\emph{Journal of Computer-Mediated Communication}} \bibinfo{volume}{25}, \bibinfo{number}{1} (\bibinfo{year}{2020}), \bibinfo{pages}{74--88}.
\newblock


\bibitem[Sundar and Kim(2019)]%
        {sundar2019machine}
\bibfield{author}{\bibinfo{person}{S~Shyam Sundar} {and} \bibinfo{person}{Jinyoung Kim}.} \bibinfo{year}{2019}\natexlab{}.
\newblock \showarticletitle{Machine heuristic: When we trust computers more than humans with our personal information}. In \bibinfo{booktitle}{\emph{Proceedings of the 2019 CHI Conference on human factors in computing systems}}. \bibinfo{pages}{1--9}.
\newblock


\bibitem[Takayama et~al\mbox{.}(2011)]%
        {takayama2011expressing}
\bibfield{author}{\bibinfo{person}{Leila Takayama}, \bibinfo{person}{Doug Dooley}, {and} \bibinfo{person}{Wendy Ju}.} \bibinfo{year}{2011}\natexlab{}.
\newblock \showarticletitle{Expressing thought: improving robot readability with animation principles}. In \bibinfo{booktitle}{\emph{Proceedings of the 6th international conference on Human-robot interaction}}. \bibinfo{pages}{69--76}.
\newblock


\bibitem[V{\"o}lkel et~al\mbox{.}(2020)]%
        {volkel2020trick}
\bibfield{author}{\bibinfo{person}{Sarah~Theres V{\"o}lkel}, \bibinfo{person}{Renate Haeuslschmid}, \bibinfo{person}{Anna Werner}, \bibinfo{person}{Heinrich Hussmann}, {and} \bibinfo{person}{Andreas Butz}.} \bibinfo{year}{2020}\natexlab{}.
\newblock \showarticletitle{How to Trick AI: Users' strategies for protecting themselves from automatic personality assessment}. In \bibinfo{booktitle}{\emph{Proceedings of the 2020 CHI conference on human factors in computing systems}}. \bibinfo{pages}{1--15}.
\newblock


\bibitem[Wang et~al\mbox{.}(2020)]%
        {wang2020sensing}
\bibfield{author}{\bibinfo{person}{Qiaosi Wang}, \bibinfo{person}{Shan Jing}, \bibinfo{person}{David Joyner}, \bibinfo{person}{Lauren Wilcox}, \bibinfo{person}{Hong Li}, \bibinfo{person}{Thomas Pl{\"o}tz}, {and} \bibinfo{person}{Betsy Disalvo}.} \bibinfo{year}{2020}\natexlab{}.
\newblock \showarticletitle{Sensing affect to empower students: Learner perspectives on affect-sensitive technology in large educational contexts}. In \bibinfo{booktitle}{\emph{Proceedings of the Seventh ACM Conference on Learning@ Scale}}. \bibinfo{pages}{63--76}.
\newblock


\bibitem[Wang et~al\mbox{.}(2021)]%
        {wang2021towards}
\bibfield{author}{\bibinfo{person}{Qiaosi Wang}, \bibinfo{person}{Koustuv Saha}, \bibinfo{person}{Eric Gregori}, \bibinfo{person}{David Joyner}, {and} \bibinfo{person}{Ashok Goel}.} \bibinfo{year}{2021}\natexlab{}.
\newblock \showarticletitle{Towards mutual theory of mind in human-ai interaction: How language reflects what students perceive about a virtual teaching assistant}. In \bibinfo{booktitle}{\emph{Proceedings of the 2021 CHI conference on human factors in computing systems}}. \bibinfo{pages}{1--14}.
\newblock


\bibitem[Wang and Yin(2021)]%
        {wang2021explanations}
\bibfield{author}{\bibinfo{person}{Xinru Wang} {and} \bibinfo{person}{Ming Yin}.} \bibinfo{year}{2021}\natexlab{}.
\newblock \showarticletitle{Are explanations helpful? a comparative study of the effects of explanations in ai-assisted decision-making}. In \bibinfo{booktitle}{\emph{26th international conference on intelligent user interfaces}}. \bibinfo{pages}{318--328}.
\newblock


\bibitem[Warshaw et~al\mbox{.}(2015)]%
        {warshaw2015can}
\bibfield{author}{\bibinfo{person}{Jeffrey Warshaw}, \bibinfo{person}{Tara Matthews}, \bibinfo{person}{Steve Whittaker}, \bibinfo{person}{Chris Kau}, \bibinfo{person}{Mateo Bengualid}, {and} \bibinfo{person}{Barton~A Smith}.} \bibinfo{year}{2015}\natexlab{}.
\newblock \showarticletitle{Can an Algorithm Know the" Real You"? Understanding People's Reactions to Hyper-personal Analytics Systems}. In \bibinfo{booktitle}{\emph{Proceedings of the 33rd annual ACM conference on human factors in computing systems}}. \bibinfo{pages}{797--806}.
\newblock


\bibitem[Xiao et~al\mbox{.}(2019)]%
        {xiao2019should}
\bibfield{author}{\bibinfo{person}{Ziang Xiao}, \bibinfo{person}{Michelle~X Zhou}, {and} \bibinfo{person}{Wat-Tat Fu}.} \bibinfo{year}{2019}\natexlab{}.
\newblock \showarticletitle{Who should be my teammates: Using a conversational agent to understand individuals and help teaming}. In \bibinfo{booktitle}{\emph{Proceedings of the 24th International Conference on Intelligent User Interfaces}}. \bibinfo{pages}{437--447}.
\newblock


\end{thebibliography}
